\documentclass[12pt]{article}
\usepackage{amsfonts,amsmath}
\usepackage{amssymb}
\usepackage{ytableau}

\newcommand{\dr}{{{\rm d}}}

\renewcommand{\theequation}{\thesection.\arabic{equation}}

\makeatletter \@addtoreset{equation}{section} \makeatother

{\vspace{3mm} }

\def\al{\alpha}

\def\*{\star}

\def\E2{\mathbf{E}}

\def\y{\mathbf{y}}
\def\v{\mathbf{v}}
\def\u{\mathbf{u}}

\def\pp{\mathrm{p}}
\def\t{\mathrm{t}}

\newcommand{\be}{\begin{equation}}
\newcommand{\ee}{\end{equation}}
\newcommand{\bee}{\begin{eqnarray}}
\newcommand{\beee}{\begin{array}}
\newcommand{\eee}{\end{eqnarray}}
\newcommand{\eeee}{\end{array}}
\newcommand{\gb}{\beta}
\newcommand{\gga}{\gamma}

\newcommand{\gd}{\delta}
\newcommand{\gl}{\lambda}
\newcommand{\gk}{\varkappa}
\newcommand{\gep}{\epsilon}

\newcommand{\gs}{\sigma}
\newcommand{\go}{\omega}

\newcommand{\nn}{\nonumber}

\newcommand{\p}{\partial}

\newcommand{\ff}{\frac}

\begin{document}
\begin{flushright}
FIAN/TD/08-2023\\
\end{flushright}

\vspace{0.5cm}
\begin{center}
{\large\bf Interaction of symmetric higher-spin gauge fields}

\vspace{1 cm}

\textbf{V.E.~Didenko and A.V.~Korybut}\\

\vspace{1 cm}

\textbf{}\textbf{}\\
 \vspace{0.5cm}
 \textit{I.E. Tamm Department of Theoretical Physics,
Lebedev Physical Institute,}\\
 \textit{ Leninsky prospect 53, 119991, Moscow, Russia }\\

\par\end{center}

\begin{center}
\vspace{0.6cm}
e-mails: didenko@lpi.ru, akoribut@gmail.com \\
\par\end{center}

\vspace{0.4cm}

\begin{abstract}
\noindent We show that the recently proposed equations for
holomorphic sector of higher-spin theory in $d=4$, also known as
chiral, can be naturally extended to describe interacting
symmetric higher-spin gauge fields in any dimension. This is
achieved with the aid of Vasiliev's off shell higher-spin algebra.
The latter contains ideal associated to traces that has to be
factored out in order to set the equations on shell. To identify
the ideal in interactions we observe the global $sp(2)$ that
underlies it to all orders. The $sp(2)$ field dependent generators
are found in closed form and appear to be remarkably simple. The
traceful higher-spin vertices are analyzed against locality and
shown to be all-order space-time spin-local in the gauge sector,
as well as spin-local in the Weyl sector. The vertices are found
manifestly in the form of curious integrals over hypersimplices.
We also extend to any $d$ the earlier observed in $d=4$
higher-spin shift symmetry known to be tightly related to
spin-locality.
\end{abstract}

\newpage
\tableofcontents

\section{Introduction}
The higher-spin (HS) interaction problem \cite{Bekaert:2022poo}
dates back to the early papers
\cite{Fronsdal:1978rb}-\cite{Fronsdal:1978vb}. Being notoriously
complicated, it still remains only partially solved. In
particular, a wealth of no go arguments
\cite{Weinberg:1964ew}-\cite{Bekaert:2010hw} (see also
\cite{Roiban:2017iqg}, \cite{Taronna:2017wbx} for a more recent
account) seemingly preclude nonlinear HS propagation on the
Minkowski space\footnote{See, however, \cite{Ponomarev:2022ryp},
\cite{Ponomarev:2022qkx} where the flat higher spins have been
recently reconsidered from the holography perspective.}. A green
light was given by the seminal results of Fradkin and Vasiliev
\cite{Fradkin:1986ka}, \cite{Fradkin:1987ks}, who addressed the
problem on the $AdS$ background and showed that some of the HS
cubic vertices carry inverse powers of cosmological constant. Ever
since the field develops rapidly and has led to many interesting
ideas and new results (see e.g.
\cite{Vasiliev:1989re}-\cite{David:2020fea} and
\cite{Bekaert:2022poo} for substantial, but still not yet a fully
comprehensive bibliography). One of the central results in the
field is the all-order HS generating equations first obtained in
$d=4$ in \cite{Vasiliev:1992av} and later on  for symmetric
bosonic fields in any $d$ in \cite{Vasiliev:2003ev}.

The AdS/CFT view proposed by Klebanov and Polyakov
\cite{Klebanov:2002ja} (see also \cite{Leigh:2003gk},
\cite{Sezgin:2003pt} for closely related papers) that suggests the
correspondence of Vasiliev's HS theory and the $O(N)$ vector model
has added a potentially testable playground for the weak-weak type
duality with HS fields in the bulk and free matter fields on the
boundary. This line of thought has been triggered by Giombi and
Yin as they directly computed the three-point correlation
functions of the boundary theory using the Vasiliev equations
\cite{Giombi:2009wh} confirming the holographic expectations. The
dual picture suggests that the bulk theory is supposedly healthy.
There are many indications in favor of this statement including
those at quantum level, see e.g.,
\cite{Giombi:2013fka}-\cite{Bae:2016rgm}.

In case the conjecture of \cite{Klebanov:2002ja} is
valid\footnote{An alternative option proposed in
\cite{Vasiliev:2012vf} is the boundary dual being a conformal HS
theory rather than vector model.} it is natural reconstructing HS
theory from e.g. a free boundary scalar. This way a partially
gauge fixed HS cubic action was successfully found in
\cite{Sleight:2016dba} and shown to be a perfect match with the
corresponding conformal field theory (CFT) independent results
from the Vasiliev theory
\cite{Sezgin:2017jgm}-\cite{Misuna:2017bjb}. At quartic order,
however, the reconstruction bumps into the locality problem
\cite{Bekaert:2015tva}, \cite{Sleight:2017pcz} that basically
leads to a lack of computational control over the Noether
procedure and to a potentially unverifiable final result for the
holographic vertex.

On the HS side the locality issue is present too, though in a
different form. Namely, Vasiliev equations are formulated in a way
which is invariant under field redefinitions, while the
reconstruction of HS vertices that they do amounts to solution of
a cohomological problem of fixing field
representatives\footnote{Interestingly, the $L_{\infty}$
structures arising within the unfolded approach in HS interactions
appear to be way more rigid than the usual Noether perturbations.
The former survive even beyond locality limitations.} in one way
or another. Already in \cite{Giombi:2009wh} the locality problem
develops at the cubic order (quadratic at the level of equations).
It has been partly circumvented by a smart analytic continuation
in spins \cite{Giombi:2009wh}, but is still shown to cause
troubles in the form of artificial infinities in
\cite{Boulanger:2015ova}. This problem was understood as the
cohomological one and solved at this order in
\cite{Vasiliev:2016xui} leading to a perfectly local vertex. At
the same time this analysis has raised a concern regarding a
systematic reconstruction of HS vertices from the Vasiliev
equations in a way compatible with locality, whatever this means.

The locality problem has proven to be a highly complicated one and
has rolled out the ongoing research mainly in $d=4$
\cite{Gelfond:2018vmi}-\cite{Didenko:2022qga} with a number of
important results\footnote{Some of the developed approaches in
this quest were successfully applied to HS theories in $d=3$
\cite{Korybut:2022kdx}.} already obtained. These include the
introduction of concepts of spin-locality and ultralocality,
locality theorems of \cite{Gelfond:2018vmi},
\cite{Vasiliev:2022med}, the manifest form of local interaction
vertices at the first few orders \cite{Didenko:2018fgx},
\cite{Gelfond:2021two} as well as the important observation of the
limiting star product in \cite{Didenko:2019xzz}. The latter
establishes spin-locality of the holomorphic sector (also known as
chiral) of the $4d$ theory to all orders \cite{Didenko:2022qga},
thus proving the locality conjecture of \cite{Gelfond:2018vmi} in
this sector. In addition, spin-locality itself seems likely to
hinge upon presence of the so-called HS shift symmetry
\cite{Didenko:2022qga}, \cite{Didenko:2022eso}; the physical
origin of which remains to be understood. Its observation was
inspired by the earlier result of \cite{Gelfond:2018vmi} known as
the structure lemma.

In proving locality of the holomorphic sector, the Vasiliev-type
equations in $d=4$ have been proposed \cite{Didenko:2022qga}.
Unlike the original Vasiliev equations, the former have no room
for both the holomorphic and antiholomorphic sectors of HS
interactions simultaneously. Nevertheless, their advantage is in
that they result in all-order spin-local vertices of the
holomorhic theory effortlessly.\footnote{In this regard it is
worth mentioning papers \cite{Sharapov:2022awp},
\cite{Sharapov:2022nps} that propose spin-local holomorphic
vertices too. The analysis relies on the homological perturbation
theory that, however, renders uncertainties of the form
$0\cdot\infty$, thus compromising consistency. As the system of
\cite{Didenko:2022qga} has no issues with consistency, it can be
used to offer validity checks of this proposal.} More importantly,
these equations can be straightforwardly generalized to describe
the interaction of symmetric HS gauge fields in any dimension
using the standard Vasiliev construction of the off shell HS
algebra \cite{Vasiliev:2003ev}, as well as to the $3d$ spinorial
HS systems.

The goal of this paper is to come up with a naturally spin-local
HS setup at least within the orders where spin-locality is already
known or expected. To this end we frame equations of
\cite{Didenko:2022qga} to (i) describe interacting bosonic HS
fields in $d$ dimensions and (ii) set the stage for further
analysis of locality. Our interest in the generating equations of
\cite{Didenko:2022qga} is motivated by their somewhat unexpected
feature to capture spin-local rather than nonlocal HS vertices in
an unforced way. Yet, the structure that controls consistency of
the proposed equations is quite puzzling on its own and extends
beyond the realm of application to holomorphic sector of the $4d$
HS theory.

The first problem (i) appears to be quite tractable as it rests on
the oscillator realization of HS algebra developed in
\cite{Vasiliev:2003ev}. This leads us to the generating equations
for the off shell, equivalently, unconstrained unfolded equations
of nonlinear HS fields. The wording 'off shell' means that the
equations actually describe no field dynamics rather provide with
a set of consistency conditions similar to the Bianchi identities
of general relativity.

We then study the off shell vertices. Interestingly, these turn
out to be spin-local even beyond the cubic approximation.
Specifically, we show that the so-called {\it canonical} choice of
physical fields gives rise to all-order ultralocality of the gauge
sector. As has been shown in \cite{Gelfond:2019tac}, ultralocality
is equivalent to the standard space-time spin locality. The Weyl
sector of the theory, where, in particular, the scalar resides is
shown to be all-order spin-local. The developed formalism brings
HS interaction vertices in their manifest form to any given order.
All these vertices exhibit a curious structure of integrals over
$n$-dimensional hypersimplices with $n$ being the order of
perturbations.

Apart from locality, HS vertices feature the shift symmetry
transformation properties. Shift symmetry acts most naturally in
the Fourier space as a certain invariance of vertices under shifts
in 'momenta' and, as it has been shown recently in
\cite{Didenko:2022eso}, goes hand in hand with locality being an
intriguing subject for future investigation. In particular, as
follows from analysis of \cite{Didenko:2022qga} and
\cite{Didenko:2022eso}, shift symmetry offers a class of field
redefinitions that respects spin-locality.

To set our system on shell a quotienting of the trace ideal of the
off shell HS algebra is required. Generally, this problem can be
as complicated as constructing HS interactions from scratch.
Indeed, to factor out the ideal one has to know its exact
deformation due to interactions. The trace ideal is known to be
generated by a certain $sp(2)$ in the case of totally symmetric
fields \cite{Vasiliev:2003ev}. In \cite{Vasiliev:2003ev} such
$sp(2)$ was found to all orders. Its very existence rests on the
properties of the so-called deformed oscillators that underlie the
original Vasiliev equations. These generate the required $sp(2)$
via bilinears.  The lack of the deformed oscillators in our case
makes on shell reduction challenging. For a reason, which is not
immediately obvious to us, the required $sp(2)$ is still there to
all orders, while the corresponding generators appear to be
remarkably simple as we find them to be linear in fields in a
closed form. This is one of the most important results of the
present work. Having manifested the structure of the trace ideal
one can systematically analyze on shell interactions
order-by-order, which we plan to carry out elsewhere.

The paper is organized as follows. In section \ref{unfo} following
\cite{Vasiliev:2003ev} we review the structure of HS unfolded
equations i.e., the spectrum of fields, higher-spin algebra and
its oscillator realization. Then, in \ref{geneq} we specify the
recently proposed in \cite{Didenko:2022qga} HS generating
equations for the case of the $d$-dimensional bosonic HS algebra
and lay down their basic properties. Section \ref{offshell} deals
with unconstrained that is off shell analysis of HS vertices. We
demonstrate the locality of the vertices, provide their manifest
form as integrals over hypersimplices, as well as reveal the shift
symmetry there. Section \ref{on-shell} is all about bringing our
generating equations on shell. Its central result is the manifest
global $sp(2)$ that generates the trace ideal along with the
explicitly given generators \eqref{t}. We conclude in section
\ref{vse}. The paper is supplemented with three appendixes.

\section{Unfolding of HS interactions}\label{unfo}
A powerful approach to HS problem is provided by the unfolded
formulation \cite{Vasiliev:1988sa} (see \cite{Misuna:2019ijn},
\cite{Misuna:2022cma} for a recent development). The spectrum of
fields suitable for description of HS symmetric fields in $d$
dimensions consists of 1-forms and 0-forms. The 1-forms
\cite{Vasiliev:2003ev}
\be\label{w}
\go^{a(s-1),\, b(n)}=\dr x^{\mu}\go^{a(s-1),\,
b(n)}_{\mu}=\ytableausetup{mathmode,
boxsize=1.25em}
\begin{ytableau}
{} & {\bullet} & {\bullet}
& {\bullet} & {} \\
{} &  {\bullet} & {\bullet} & {}
\end{ytableau}\,\,,\qquad 0\leq n\leq
s-1
\ee
are given by the $o(d-1,1)$ two-row Young diagrams
\be\label{youngw}
\go^{a(s-1),\, ab(n-1)}=0\,,\qquad a,b=0,\dots d-1\,,
\ee
where as typical in HS literature we denote by one and the same
letter symmetrization over a group of indices with eq.
\eqref{youngw} being so the standard Young condition. At free
level one imposes the irreducible traceless condition
\be\label{wtr}
\go^{a(s-3)c}{}_{c,}{}^{b(n)}=0\,.
\ee
The physical component which contains the usual Fronsdal spin-$s$
field corresponds to $n=0$, while $n>0$ are auxiliary, i.e., they
can be expressed in terms of derivatives of the physical field.
Note that for each $s$ there are finitely many $\go$'s. Thus
introduced $\go$'s play the role of HS potentials generalizing
Maxwell potential $A=\go$ for $s=1$ and the Lorentz connection
$\go^{a,b}$ along with the frame field $\mathbf{e}^{a}=\go^a$ of
the Cartan gravity for $s=2$.

The 0-forms
\be\label{C}
C^{a(m),\, b(s)}=\ytableausetup{mathmode,
boxsize=1.25em}
\begin{ytableau}
{} & {\bullet} & {\bullet}
& {\bullet} & {} \\
{} &  {\bullet} & {\bullet} & {}
\end{ytableau}\,\,,\qquad m\geq s
\ee
having an arbitrary long first row are unbounded for fixed $s$.
The $m=s$ component corresponds to the rectangular HS Weyl tensors
generalizing Maxwell tensor $F^{a,b}=C^{a,b}$ for $s=1$ and the
Weyl tensor $C^{ab,cd}$ for $s=2$. Components $m>s$ are auxiliary,
being expressed roughly via derivatives of the primary fields with
$m=s$. Similarly, $C$ is subject to the Young constraint
\be\label{youngC}
C^{a(m),\, ab(s-1)}=0
\ee
along with the traceless condition
\be\label{Ctr}
C^{a(m-2)c}{}_{c,}{}^{b(s)}=0\,.
\ee
A convenient way to work with the two-row Young tensors was
proposed in \cite{Vasiliev:2003ev}. Accordingly, let us introduce
the following commuting variables $\vec\y_{\al}:=\y^{a}_{\al}$ and
$y_{\al}$, where $\al=1,2$. The two-component indices can be
associated with an $sp(2)$ algebra. To this end one introduces the
canonical $sp(2)$ form $\gep_{\al\gb}=-\gep_{\gb\al}$ and its
inverse $\gep^{\al\gb}\gep_{\al\gga}=\gd_{\gga}{}^{\gb}$ and
defines the lowering and raising index convention, e.g.,
\be
y^{\al}=\gep^{\al\gb}y_{\gb}\,,\qquad
y_{\al}=y^{\gb}\gep_{\gb\al}\,,
\ee
then the Taylor coefficients of any analytic function $f(\vec\y,
y)$ are supplemented with the following condition:
\be\label{spop}
\left(\vec\y_{\al}\cdot\ff{\p}{\p \vec\y^{\gb}}+y_{\al}\ff{\p}{\p
y^{\gb}}+\al\leftrightarrow\gb\right)f=0\,,
\ee
which forms the two-row Young diagrams with respect to Lorentz
indices $a,b,\dots$. In other words, they satisfy conditions like
\eqref{youngw}, but not the traceless constraint \eqref{wtr},
which has to be imposed additionally in order to stay with
$o(d-1,1)$ irreducible components. Eq. \eqref{spop} has the
transparent meaning of an $sp(2)$ invariance. Indeed, by
introducing the standard Moyal star product
\be\label{moyal}
(f*g)(y, \vec\y)= \int f\left(y+u, \vec\y+\vec\u\right)
g\left(y+v, \vec\y+\vec\v\right)
e^{iu_{\al}v^{\al}+i\vec\u_{\al}\vec\v^{\al}}\,,
\ee
where integration over $u$, $v$, $\vec{\mathbf{u}}$ and $\vec{\mathbf{v}}$ is assumed, it yields the
following commutation relations
\be
[\y^{a}_{\al}, \y^{b}_{\gb}]_*=2i\eta^{ab}\gep_{\al\gb}\,,\qquad
[y_{\al}, y_{\gb}]_*=2i\gep_{\al\gb}\,,\qquad [\y^{a}_{\al},
y_{\gb}]_*=0\,.
\ee
One then finds that \eqref{spop} can be written down as
\be\label{singlet}
[t_{\al\gb}^0,f(y, \vec\y)]_*=0\,,\qquad
t_{\al\gb}^0=\ff{1}{4i}(\y^{a}_{\al}\y_{a\gb}+y_{\al}y_{\gb})\,.
\ee
Generators $t^0_{\al\gb}$ form the so-called Howe dual algebra to
$o(d-1,2)$ spanned by various $sp(2)$-invariant bilinears of
$\vec\y$ and $y$
\be
P^{a}=\y^{a}_{\al}y^{\al}\,,\qquad M^{ab}=\y^{a}_{\al}\y^{b\al}\,.
\ee
It is easy to observe $sp(2)$ commutation relations for the
generators defined in \eqref{singlet}
\be\label{sp20}
[t^0_{\al\gb},
t^{0}_{\gga\gd}]_*=\gep_{\al\gga}t_{\gb\gd}^0+\gep_{\gb\gga}t_{\al\gd}^0+
\gep_{\al\gd}t_{\gb\gga}^0+ \gep_{\gb\gd}t_{\al\gga}^0\,.
\ee

Having introduced the above formalism, the HS algebra can be
constructed as follows \cite{Vasiliev:2004cm}. Consider an
associative algebra spanned by various formal power series in
$\vec{\mathbf{y}}$ and $y$ that has product \eqref{moyal}. This
algebra contains a subalgebra $\mathcal{S}$ generated by elements
that fulfil $sp(2)$ singlet condition \eqref{singlet}. Algebra
$\mathcal{S}$ is not simple since it contains two-sided ideal $I$
of the form
\be\label{id}
g=t^{0}_{\al\gb}*f^{\al\gb}(y, \vec\y)=f^{\al\gb}(y,
\vec\y)*t^{0}_{\al\gb}\in I\,,
\ee
where $g$ satisfies \eqref{singlet} which in practice means that
$sp(2)$ indices of $f_{\al\gb}$ should be carried by $y_{\al}$ and
$\vec\y_{\al}$. One can consider the quotient algebra
$\mathcal{S}/I$ which is still an associative algebra and then
build out of it a Lie algebra by using the Lie bracket as a
commutator. The latter algebra is the HS algebra in $d$
dimensions.\footnote{This algebra was originally obtained in terms
of the enveloping algebra in \cite{Eastwood:2002su}, where higher
symmetries of the massless scalar were studied. The corresponding
algebra was called conformal higher spin algebra.} Treating
algebra $\mathcal{S}$ as a linear space one can strip its ideal
off. This means the following decomposition for general $f\in
\mathcal{S}$
\be\label{sp2free}
f=\mathbf{f}+g\,,\;\;\; g\in I\,,
\ee
where one has to specify any particular way of choosing the
representative. For the $\mathbf{f}\in \mathcal{S}/I$ choice, its
coefficients in Taylor expansion can be chosen to correspond to
traceless Young diagrams [e.g., \eqref{wtr}].

Following the standard HS philosophy, consider now the 1-form
$\go$ and 0-form $C$ taking values in the off shell HS algebra.
These encode fields \eqref{w} and \eqref{C} in the Taylor-like
expansion. The unfolded equations are
\begin{align}
&\dr_x\go+\go*\go=\Upsilon(\go,\go, C)+\Upsilon(\go,\go, C,
C)+\dots\,,\label{eqw}\\
&\dr_x C+\go*C-C*\pi(\go)=\Upsilon(\go, C, C)+\Upsilon(\go, C, C,
C)+\dots\,,\label{eqC}
\end{align}
where $\pi$ is the reflection automorphism
\be
\pi f(y, \vec\y)=f(-y, \vec\y)
\ee
and $\Upsilon$'s are interaction terms that describe the HS
gauge-invariant off shell field identities. Abusing the
terminology we call such vertices 'off shell' as opposed to the on
shell ones based on the factor algebra corresponding to dynamical
HS interaction. To arrive at the on-shell HS evolution one has to
strip off the associated $sp(2)$ generated ideal. As mentioned,
this problem can be really complicated due to the fact that
$\Upsilon$'s on the right-hand side of \eqref{eqw} and \eqref{eqC}
are not built out of field star products. Instead, the off shell
HS algebra gets deformed which entails the deformation of the
ideal. As a result the factorization cannot be carried out using
the undeformed prescription \eqref{sp2free}. In other words,
generators of the $sp(2)$ algebra defined in \eqref{singlet}
become field dependent in the interactions.

\subsection{Generating equations}\label{geneq}

To arrive at eqs. \eqref{eqw}-\eqref{eqC} we follow
\cite{Vasiliev:2003ev} and \cite{Didenko:2022qga}. Namely, using
Vasiliev's trick we enhance the space spanned by $Y=(y, \vec\y)$
by introducing the auxiliary two-component\footnote{Notice,
however, that we do not double variables $Y$'s unlike
\cite{Vasiliev:2003ev}. Our choice corresponds to a particular
gauge fixing of St\"{u}ckelberg fields along with the conventional
choice for the compensator.} $z_{\al}$ and embed the 1-form $\go$
into this larger space
\be\label{embd}
\go\to W(z; Y):=\go(Y)+W_1(z;Y)+W_{2}(z;Y)+\dots\,,
\ee
where $W_n$ are yet to be defined perturbative in $C$ corrections. We call embedding \eqref{embd} {\it canonical} provided
\be\label{canembd}
W_{n}(0, Y)=0\,,\qquad\forall n\geq 1
\ee
which means that the physical field representative $\go=W(0, Y)$
is chosen to be fixed and is not subject to field redefinition in
the interactions; this would not be the case if $W_k(0, Y)\neq 0$
for some integer $k$. To avoid any confusion at this point we want
to stress here that constraint \eqref{canembd} just defines the
way one can solve the generating system rather than forbids field
redefinitions.\footnote{Similarly, the so-called $z$-dominated
constraints have been imposed (see e.g. \cite{Gelfond:2018vmi}) to
the first few orders in solving Vasiliev's generating system to
obtain spin-local vertices.}

We then also enhance the Moyal star product \eqref{moyal} to a
large $z$-dependent product \cite{Didenko:2022qga} (see also
\cite{Didenko:2019xzz})
\be\label{limst}
(f*g)(z; Y)= \int f\left(z+u', y+u; \y \right)\star
g\left(z-v,y+v+v'; \y \right)
\exp({iu_{\al}v^{\al}+iu'_{\al}v'^{\al}})\,,
\ee
where $\star$ is a part of star product \eqref{moyal} that acts on
$\vec\y$ only
\be\label{starvec}
(f\star g)(\vec\y)= \int f\left(\vec\y+\vec{\mathbf{u}}\right)
g\left(\vec\y+\vec{\mathbf{v}} \right)
\exp({i\vec{\mathbf{u}}_{\al}\vec{\mathbf{v}}^{\al}})\,.
\ee
Notice, the $(z,y)$ ordering in \eqref{limst} differs from the
original Vasiliev one \cite{Vasiliev:2003ev}. For $z$-independent
functions \eqref{limst} is identical to \eqref{moyal}. Moreover,
\eqref{limst} coincides with that of Vasiliev if either $f$ or $g$
is $z$-independent. The large star product mixes $z$ and $y$
leaving transformation with respect to $\vec\y$ unaffected. A
distinguishing property of \eqref{limst} is
\be
[z_{\al}, z_{\gb}]_*=0\,.
\ee
More generally, one comes along with the following action
\begin{align}
&y* =y+i\ff{\p}{\p y}-i\ff{\p}{\p z}\,,\qquad z* =z+i\ff{\p}{\p
y}\,,\\
&* y=y-i\ff{\p}{\p y}-i\ff{\p}{\p z}\,,\qquad * z=z+i\ff{\p}{\p
y}\,,\\
&\vec\y* =\vec\y+i\ff{\p}{\p \vec\y}\,,\qquad *
\vec\y=\vec\y-i\ff{\p}{\p \vec\y}\,.
\end{align}
The important element of the star product \eqref{limst} is a
``$\gd$-function''
\be\label{delta}
\gd:=e^{iz_{\al}y^{\al}}\,,\qquad f(z)*\gd=\gd*f(z)=\gd\cdot
f(0)\,.
\ee
Its square $\gd*\gd=0\cdot\infty$ is ill-defined that along with
\eqref{delta} emphasizes an analogy with
$\gd$-function.\footnote{Let us stress that the square of
$\gk=\exp{izy}$ is perfectly well-defined in the Vasiliev
ordering, $\gk*\gk=1$, \cite{Vasiliev:2003ev}.}

Generating equations that reconstruct right-hand sides of
\eqref{eqw} and \eqref{eqC} order-by-order has the form
\cite{Didenko:2022qga}
\begin{align}
&\dr_x W+W*W=0\,,\label{xeq}\\
&\dr_z W+\{W,\Lambda\}_*+\dr_x\Lambda=0\,,\label{zeq}\\
&\dr_x C+\left(W(z'; y, \vec\y)*C-C*W(z'; -y,
\vec\y)\right)\Big|_{z'=-y}=0\,,\label{dCeq}
\end{align}
where $C=C(y, \vec\y |\, x)$ is $z$-independent and $\Lambda$ is
the following $\dr z$-form
\be\label{lambda}
\Lambda=\dr z^{\al}z_{\al}\int_{0}^{1}\dr\tau\,\tau C(-\tau z,
\vec\y)e^{i\tau z_{\al}y^{\al}}\,,
\ee
which satisfies
\be\label{dzlambda}
\dr_z\Lambda=C*\gga\,,\qquad \gga=\ff12\, e^{izy}\,\dr z^{\gb}\dr
z_{\gb}\,\;\;\;\; \dr_z:=\dr z^\alpha \frac{\partial}{\partial
z^\alpha}\,,
\ee
where we use the following short hand notation for index
contraction
\be\label{contr}
\xi_{\al}\eta^{\al}:=\xi\eta=-\eta\xi\,.
\ee
Note that eq. \eqref{dCeq} unlike \eqref{xeq} and \eqref{zeq}
contains no $z$ and, therefore the star product acts in eq.
\eqref{dCeq} by ignoring dummy  argument $z'$ of $W$ which is set
to $z'=-y$ at the end.

\paragraph{Consistency.} Unlike the original Vasiliev equations,
consistency of system \eqref{xeq}-\eqref{dCeq} is far from being
obvious. It rests on the properties of functional class that
admits evolution on \eqref{xeq}-\eqref{dCeq} and a particular {\it
projective identity} \eqref{iden}. All necessary details are laid
in \cite{Didenko:2022qga}. Here we briefly sketch the basic steps:
\begin{itemize}
\item First, apart from formal consistency it is important to
examine whether or not star products of various master fields in
\eqref{xeq}-\eqref{dCeq} exist at all. For example,
$\Lambda*\Lambda$ can be checked to be ill-defined. To answer that
question one searches for a class of functions that respects
operations on the generating equations. It turns out that if
system \eqref{xeq}-\eqref{dCeq} has perturbative solutions, then
they belong to $\dr z$-graded class ${\mathbf{C}}^r$, where $r$ is
the rank of $\dr z$-differential form (see Appendix A),
\begin{align}
&{\mathbf{C}}^{r_1}*{\mathbf{C}}^{r_2}\to{\mathbf{C}}^{r_1+r_2}\,,\quad r_1+r_2<2\label{C1}\\
&\dr_z {\mathbf{C}}^{r}\to{\mathbf{C}}^{r+1}\,,\qquad \dr_x
{\mathbf{C}}^{r}\to{\mathbf{C}}^{r}\,.\label{C2}
\end{align}
Correspondingly, $W\in {\mathbf{C}}^{0}$, $\Lambda\in
{\mathbf{C}}^{1}$ and $C\in {\mathbf{C}}^{0}$. In
\cite{Didenko:2022qga} it was proven that functions \eqref{class}
indeed respect requirements \eqref{C1}-\eqref{C2} and result in
well-defined star products provided $r_1+r_2\neq 2$. The following
products ${\mathbf{C}}^{0}*{\mathbf{C}}^{2}$,
${\mathbf{C}}^{2}*{\mathbf{C}}^{0}$ and
${\mathbf{C}}^{1}*{\mathbf{C}}^{1}$ are sick unless
${\mathbf{C}}^{0}$ is $z$-independent. Luckily, there are no
ill-defined products in \eqref{xeq}-\eqref{dCeq}. It was also
shown that arbitrary $z$-independent functions of  grading $r=0$
do belong to ${\mathbf{C}}^{0}$, while $\gd\in {\mathbf{C}}^{2}$
\eqref{delta}.

\item Now that we have identified proper functions \eqref{class},
one proves the following projecting identity
\cite{Didenko:2022qga}. Suppose $\phi(z,y; \vec\y)\in
{\mathbf{C}}^{0}$ is a $\dr x$ 0-form,\footnote{For space-time
differential forms of higher ranks one should bear in mind $\{\dr
x, \dr z_\al\}=0$ that may result in overall signs in
\eqref{iden}} then
\be\label{iden}
\dr_z (\phi*\Lambda)=(\phi(z',y;
\vec\y)*C)\Big|_{z'=-y}*\gga\,,\qquad \dr_z
(\Lambda*\phi)=(C*\phi(z',-y; \vec\y))\Big|_{z'=-y}*\gga\,.
\ee
In proving \eqref{iden} one needs \eqref{class} and
\eqref{lambda}. Identity \eqref{iden} was coined {\it projective}
in \cite{Didenko:2022qga}, since it somehow projected out the
dependence on $z$ of $\phi$ into $\gd$-element \eqref{delta}
stored in $\gga$. We note also, that while it is tempting to use
the Leibniz rule and eq. \eqref{dzlambda} in proving \eqref{iden},
this would lead to uncertainty because both terms in decomposition
$\dr_z({\mathbf{C}}^{0}*{\mathbf{C}}^{1})=\dr_z
{\mathbf{C}}^{0}*{\mathbf{C}}^{1}+
{\mathbf{C}}^{0}*\dr_z{\mathbf{C}}^{1}$ do not exist
separately.\footnote{It is worth to mention that the failure of
Leibniz rule offers some freedom in definition of $\dr_z$ in
sector of $\dr z$ one-forms in a way that may differ from
\eqref{dzlambda}. Unfortunate choice might lead to inconsistent
vertices, however the one used in this paper and in
\cite{Didenko:2022qga} is consistent. More details are in
\cite{Korybut_in_prep}.}

\item Armed with the class \eqref{C1} and \eqref{C2} and identity
\eqref{iden} all is set to guarantee consistency of
\eqref{xeq}-\eqref{dCeq}. Equation \eqref{xeq} is clearly
consistent under $\dr_x^2=0$. Hitting \eqref{xeq} with $\dr_x$ and
substituting $\dr_x W$ from \eqref{xeq} back one faces no
ill-defined objects either. Now, applying $\dr_z$ to \eqref{xeq}
and using \eqref{zeq} we again find neither new constraints nor
ill-defined expressions. Less trivial is to check consistency of
\eqref{zeq} and \eqref{dCeq}. Applying $\dr_z$ to \eqref{zeq}
gives the following constraint
\be\label{cstr}
\dr_z\{W,\Lambda\}_*=\dr_x C*\gga\,.
\ee
The problem here is that one can not use Leibniz rule on the left
of \eqref{cstr} because product of
${\mathbf{C}}^{r_1}*{\mathbf{C}}^{r_2}$ with $r_1+r_2=2$ is
ill-defined as noted earlier. Another point is since $\dr_x C$ is
$z$-independent by definition, the $z$-dependence of the left-hand
side of \eqref{cstr} must be of the very specific form
$F(Y)*\gga$. At this stage identity \eqref{iden} helps. Using it
one has
\be
\left(C*W(z'; -y, \vec\y)-W(z'; y,
\vec\y)*C\right)\Big|_{z'=-y}*\gga=\dr_x C*\gga\,,
\ee
which holds in view of \eqref{dCeq}. Finally, one has to check
that \eqref{dCeq} is consistent with $\dr_x^2=0$. This is indeed
so, because \eqref{dCeq} comes out as a consequence of \eqref{zeq}
in the form of \eqref{cstr}, which is equivalent to \eqref{dCeq}.
The former is manifestly consistent as can be seen upon applying
$\dr_x$ to it and using again \eqref{xeq}, \eqref{zeq}. Let us
also note in this regard that \eqref{dCeq}, therefore,  is not an
independent equation.
\end{itemize}

\paragraph{Gauge symmetries} Equations
\eqref{xeq}-\eqref{dCeq} are invariant under the following local
gauge transformations parameterized by $\gep\in {\mathbf{C}}^{0}$
\begin{align}
&\gd_\gep W=\dr_x\gep+[W,\gep]_*\label{locW}\\
&\gd_\gep\Lambda=\dr_z\gep+[\Lambda, \gep]_*\label{locL}\,,\\
&\gd_\gep C=(\gep(z',y)*C-C*\gep(z',-y))_{z'=-y}\,.\label{locC}
\end{align}
The above transformations leave \eqref{xeq}-\eqref{dCeq} and
\eqref{dzlambda} invariant, but not yet \eqref{lambda}. In order
to keep \eqref{lambda} intact one has to require
\be\label{deltaL}
\gd_{\gep}\Lambda(C)=\Lambda(\gd_\gep C)\,,
\ee
which is equivalent to
\be\label{true}
\dr_{z}\gep+[\Lambda(C), \gep]_*-\Lambda(\gd_\gep C)=0\,,
\ee
where the last term in \eqref{true} is understood as
\eqref{lambda} with $C$ being replaced by \eqref{locC}. Equation
\eqref{true} is consistent as can be seen upon acting with $\dr_z$
and using \eqref{iden} and, therefore, can be solved
order-by-order using the standard contracting homotopy. Functions
$\gep\in\mathbf{C}^0$ that satisfy \eqref{true} determine true
local gauge symmetries of HS generating equations
\eqref{xeq}-\eqref{dCeq}. The corresponding solution space we
label by $\mathbf{C}^0_{c}$, where subscript $C$ stands for a
$z$-independent element that $\Lambda$ depends on. It is easy to
see that these functions form the Lie algebra. Indeed, on one hand
from \eqref{locL} one finds
\be\label{comL}
(\gd_\gep\gd_\eta-\gd_\eta\gd_\gep)\Lambda=\dr_z[\gep,
\eta]_*+[\Lambda, [\gep, \eta]_*]_*\,.
\ee
On the other hand, from \eqref{deltaL} it follows
$\gd_{\eta}\gd_{\gep}\Lambda=\Lambda(\gd_{\eta}\gd_{\gep} C)$ and
so
\be
(\gd_\gep\gd_\eta-\gd_\eta\gd_\gep)\Lambda=\Lambda((\gd_\gep\gd_\eta-\gd_\eta\gd_\gep)C)\,,
\ee
where we have also taken into account that $\Lambda$ is linear in
$C$
\be
\Lambda(C_1+C_2)=\Lambda(C_1)+\Lambda(C_2)\,.
\ee
Those parameters $\gep^{gl}\in\mathbf{C}^0_{c}$ that leave fields
$W$, $\Lambda$ and $C$ invariant generate global symmetries
\be
\gd W=\gd\Lambda=\gd C=0\,.
\ee
In this case the corresponding global symmetry parameter satisfies
\begin{align}
&\dr_x\gep^{gl}+[W, \gep^{gl}]_*=0\,,\\
&\dr_z\gep^{gl}+[\Lambda, \gep^{gl}]_*=0\,.\label{glz}
\end{align}
Note that $\gd_{\gep^{gl}}C=0$ comes about as the integrability
consequence of \eqref{glz}. It also entails validity of
\eqref{true}.

\paragraph{Projectively twisted-adjoint module.} The role of
projective identity \eqref{iden} is crucial for consistency
\eqref{xeq}-\eqref{dCeq}. One of its remarkable properties is it
generates a kind of a twisted-adjoint representation of the large
algebra that acts on its module spanned by $z$-independent
functions from the off shell HS algebra. Since the module is
$z$-independent, we call it {\it projective}. Specifically, for a
$z$-independent element $C(y, \vec\y)$ and any $\gep\in
{\mathbf{C}}^{0}_{c}$ let us define the following action
\be\label{rho}
\rho_\gep(C):=(\gep(z',y)*C-C*\gep(z',-y))_{z'=-y}\,.
\ee
Recall, that star product in \eqref{rho} is insensitive to $z'$,
which stands idle there and then is set to $-y$. One can show that
$\rho$ forms a representation of the large Lie algebra, as for two
such $\gep_{1,2}$ we have
\be\label{rep}
(\rho_{\gep_1}\rho_{\gep_2}-\rho_{\gep_2}\rho_{\gep_1})C=\rho_{[\gep_1,
\gep_2]_*}C\,.
\ee
The proof of this fact follows from \eqref{comL}. A brief comment
in regard of \eqref{rep} is in order. We can not make a stronger
statement than \eqref{rep}. Looking at the large algebra as at the
associative one, we could have expected the following map
\be
\tau_{\gep}(C):=(\gep(z',y)*C)_{z'=-y}
\ee
to be its representation. However, this is not the case as it is
not difficult to show that
\be
\tau_{\gep_1}\tau_{\gep_2}\neq \tau_{\gep_1*\gep_2}
\ee
for $z$-dependent $\gep_{1,2}\in {\mathbf{C}}^{0}$.

Notice, \eqref{rep} guarantees that local gauge symmetries act
properly on the Weyl module $C$ to form an algebra, i.e.,
$\gd_{[\gep_1}\gd_{\gep_2]}=\gd_{[\gep_1, \gep_2]_*}$. It is
important also to stress that parameter $\gep(z, Y|x)$ is {\it
not} an arbitrary analytic function of $z$ and $Y$ but belongs to
a much smaller subset of analytic functions from
${\mathbf{C}}^{0}$. Still, the latter class is big enough to
capture HS dynamics as it contains all analytic $z$-independent
functions.

\section{Off shell: vertices, locality and shift symmetry}\label{offshell}

Equations \eqref{xeq}-\eqref{dCeq} generate unconstrained, which
is off shell, HS vertices of \eqref{eqw} and \eqref{eqC}. The
procedure is pretty standard for the Vasiliev-like systems. In our
case it boils down to a resolution for $z$-dependence of field $W$
using \eqref{zeq} and substitution of the result into \eqref{xeq}
and \eqref{dCeq}. Equation \eqref{dCeq} being manifestly
$z$-independent reproduces \eqref{eqC}. While $z$-independence of
\eqref{xeq} that brings \eqref{eqw} is not manifest, it is
guaranteed by the fact that application of $\dr_z$ to the
left-hand side of \eqref{xeq} is zero provided \eqref{zeq}
imposed. This makes the analysis particularly simple as
substitution of any $W$ that satisfies \eqref{zeq} into
\eqref{xeq} can be carried out at an arbitrary convenient value of
$z$, e.g., at $z=0$. Below we provide all necessary details. The
important conclusion here is the off shell vertices appear to be
properly local at any interaction order. Namely, we will show that
all $\Upsilon(\go, \go, C\dots C)$ are space-time spin-local,
while all $\Upsilon(\go, C\dots C)$ are spin-local.

Another observation, which is tightly related to locality is the
shift symmetry of the obtained vertices. It has been already
observed in \cite{Didenko:2022qga} as a symmetry of the
holomorphic on shell HS vertices in four dimensions. Its effect
has been analyzed on general grounds within the Vasiliev theory in
$d=4$, \cite{Didenko:2022eso} with a conclusion that if present,
it leads to HS spin-locality under some mild extra assumptions.

\paragraph{Contracting homotopy} First order in $z$
partial differential equation \eqref{zeq} determines
$z$-dependence of the perturbative corrections $W_{n}(z; Y)$ from
\eqref{embd}. It brings a freedom in $z$-independent functions at
each perturbation order. This freedom is naturally interpreted as
field redefinition of the physical field $\go(Y)$ by higher order
in $C$ terms
\be
\go\to \go+F(\go, C\dots C)\,.
\ee
Field $\go$ makes its appearance at zeroth order. For reasons that
will become significant upon on-shell reduction of HS equations,
we stick to the canonical form \eqref{canembd} of perturbative
corrections $W_n$ implying that we set
\be
F(\go, C\dots C)=0\,.
\ee
In other words, we fix field $\go$ at the very beginning assuming
no further field redefinitions. This choice lives up to the
standard contracting homotopy as will be clear soon.

Consider the following equation for $\dr z$-zero form $f(z)$
\be
\dr_z f(z)=g(z):=\dr z^{\al}g_{\al}(z)\,.
\ee
It is consistent provided $\dr_z g(z)=0$. In this case it can be
solved up to a constant as
\be\label{hom}
f(z)=\Delta g(z):=z^{\al}\int_{0}^{1}\dr\tau\,g_{\al}(\tau z)\,.
\ee
Operator $\Delta$ is referred to as the contracting homotopy
operator. Note, that no part of $g_{\al}(z)$ of the form
$z_{\al}\phi(z)$ contributes to the solution for $f$, since
$\Delta(\dr z^{\al} z_{\al}\phi(z))\equiv 0$ as
$z^{\al}z_{\al}=0$. Now, substituting decomposition \eqref{embd}
into \eqref{zeq} one finds
\be
\dr_z W_n=-\{W_{n-1}, \Lambda\}_*-(\dr_x\Lambda)_n\,,\qquad
W_0:=\go(Y|x)\,.
\ee
The solution satisfying \eqref{canembd} can be then written down
as
\be\label{Wsol}
W_n=-\Delta \{W_{n-1}, \Lambda\}_*\,,
\ee
where $(\dr_x\Lambda)_n$, being proportional to $z_{\al}$,
\eqref{lambda} vanishes upon application of $\Delta$. Starting
from physical $\go$, eq. \eqref{Wsol} generates any order $O(C^n)$
corrections $W_n(z; Y)$ such, that $W_n(0, Y)=0$ in accordance
with the canonical embedding \eqref{canembd}. The corresponding HS
vertices from \eqref{eqw} and \eqref{eqC} acquire the following
form
\begin{align}
&\Upsilon(\go, \go, \underbrace{C\dots
C}_n)=-\left(\sum_{j+k=n}W_j*W_k\right)\Big|_{z=0}\,,\label{wwver}\\
&\Upsilon(\go, \underbrace{C\dots C}_n)=-\left(W_{n-1}(z'; y,
\vec\y)*C-C*W_{n-1}(z'; -y,
\vec\y)\right)\Big|_{z'=-y}\,,\label{wver}
\end{align}
where in \eqref{wwver} we set $z=0$ for convenience as the vertex
is $z$-independent anyway. Since $W_n(0; Y)=0$ for $n>0$, this
choice leads to no contribution from $\dr_x W_n$.

\subsection{Vertices}
Star product \eqref{limst} is well-suited for exponentials. It is
then convenient to use the Taylor representation with respect to
variable $y$ of our fields
\begin{align}
&C(y, \vec\y) \equiv e^{-i\pp^{\al}y_{\al}}
C(y',\vec\y)\Big|_{y'=0}\,,
\quad \pp_{\al}=-i\ff{\p}{\p y^{'\al}}\,,\label{conv1}\\
&\go(y, \vec\y)\equiv
e^{-i\t^{\al}y_{\al}}\go(y',\vec\y)\Big|_{y'=0}\,,\quad
\t_{\al}=-i\ff{\p}{\p y^{'\al}}\,,\label{conv2}
\end{align}
where $\pp$ acts on $C$, while $\t$ on $\go$. As the HS vertices,
$\Upsilon$, from \eqref{eqw} and \eqref{eqC} are strings of
several $C$'s the number of which depends on the order of
perturbations, we endow $\pp$ with index $\pp_k$ that points at
which $C$ it acts upon as seen from left. Similarly, in
\eqref{eqw} where two $\go$'s present, we distinguish between
$\t_1$ and $\t_2$. The only star product that one faces in
extracting vertices is the one of the gaussian exponentials.
Notice also, that $\star$-product with respect to $\vec\y$'s stays
undeformed in interaction that makes it dummy in practical
calculation. Therefore, the vertices can be conveniently written
down in the generating form using functions $\Phi^{[\delta_1,\,
\delta_2]}(y; \t_{1,2}, \pp_1,\dots, \pp_n)$ that reproduce
various order $n$ vertex contributions via
\begin{align}
&\Upsilon=\sum_{\gd_2>\gd_1}\Phi^{[\delta_1,\, \delta_2]}(y; \t_1,
\t_2, \pp_i)\left(C{\star}\dots{\star}\,\go{\star}\dots{\star}\,
\go{\star}\dots{\star}\,C\right)(y'_{I}, \vec\y)\Big|_{y'_{I}=0}
\,,\label{ups1}
\end{align}
where $\gd_{1,2}$ labels positions of two $\go$'s in the string of
\eqref{ups1}. Vertices from \eqref{eqC} look similar
\begin{align}
&\Upsilon=\sum_{\gd}\Phi^{[\delta]}(y; \t,
\pp_i)\left(C{\star}\dots{\star}\,\go{\star}\dots{\star}\,C\right)(y'_{I},
\vec\y)\Big|_{y'_{I}=0}\,.\label{ups2}
\end{align}
So, the HS vertices are unambiguously determined via functions
$\Phi$, which we will call the off shell or unconstrained vertices
abusing the terminology. We will also ignore the part on which
operators $\t$ and $\pp$ act upon, that is we adopt the following
shorthand notation
\begin{align}
&\go\to e^{-i\t^{\al}y_{\al}}\,,\label{wexp}\\
&C\to e^{-i\pp^{\al}y_{\al}}\,,\label{Cexp}\\
&\Lambda\to\dr z^{\al}\int_{0}^{1}\dr\tau\,\tau z_{\al}e^{i\tau
z_{\gb}(y+\pp)^{\gb}}\,.\label{Lexp}
\end{align}
Using \eqref{Wsol} it is not difficult to arrive at the manifest
expressions for $\Phi$ at any order. However, since these vertices
are mere the generalized Bianchi constraints, one has to strip off
the traceful ideal properly to set them on shell. This problem
will be addressed elsewhere, while here we provide with examples
of all order off shell vertices, as well as spell out some
properties that in fact do not rely on their explicit form.

Using that $W_0=\go(y, \vec\y)$ at the lowest order one finds from
\eqref{Wsol} and \eqref{wexp}, \eqref{Lexp}
\be
W_1=W^{(1)}_{\go C}+W^{(1)}_{C\go}\,,\qquad W^{(1)}_{\go
C}=-\Delta(\go*\Lambda)\,,\qquad
W^{(1)}_{C\go}=-\Delta(\Lambda*\go)\,,
\ee
\begin{align}
&W^{(1)}_{\go C}=\t^{\al}z_{\al}\int\dr^{3}_{\triangle}\tau\,
e^{i\tau_1 z_{\al}(y+\pp+\t)^{\al}+i\tau_2
\t_{\al}y^{\al}+i(1-\tau_2)\pp_{\al}t^{\al}}\,,\\
&W^{(1)}_{C\go}=-\t^{\al}z_{\al}\int\dr^{3}_{\triangle}\tau\,
e^{i\tau_1 z_{\al}(y+\pp-\t)^{\al}+i\tau_2
\t_{\al}y^{\al}+i(1-\tau_2)\pp_{\al}\t^{\al}}\,,
\end{align}
where we have used the following shorthand notation for
$\tau$-integrals
\begin{equation}
\int \dr^3_\triangle\tau :=\int \dr \tau_1 \, \dr \tau_2 \dr \tau_3\, \theta(\tau_1)\theta(\tau_2)\theta(\tau_3)\, \delta(1-\tau_1-\tau_2-\tau_3)\,.
\end{equation}
This simplex integration arises upon suitable integration variable
change along the lines of \cite{Vasiliev:2016xui},
\cite{Didenko:2018fgx}. The vertex is then obtained via
\eqref{wwver}
\be
\Upsilon(\go, \go, C)=-(\go*W_1+W_1*\go)_{z=0}
\ee
with the final result readily accessible
\begin{align}
&\Phi^{[1,2]}=\t_1 \t_2 \int\dr^3_{\triangle}\tau\,
e^{i(1-\tau_1)\,\t_1 {}_\alpha y^\alpha+i\tau_2\, \t_2 {}_\alpha
y^\alpha-i(\tau_1 t_1+(1-\tau_2)\, \t_2)_\alpha\pp^\alpha
+i(1-\tau_3)\t_{2\alpha} \t_1^\alpha}\,,\label{wwC}\\
&\Phi^{[1,3]}=\t_2 \t_1
\int\dr^3_{\triangle}\tau\,e^{i(1-\tau_1)\,\t_2 {}_\alpha
y^\alpha+i\tau_2\, \t_1 {}_\alpha y^\alpha-i(\tau_1
\t_2+(1-\tau_2)\,\t_1)_\alpha \pp^\alpha
+i(\tau_2-\tau_1)\t_2 {}_\alpha \t_1^\alpha }+\label{wCw}\\
&+\t_2\t_1 \int\dr^3_{\triangle}\tau\,e^{i(1-\tau_1)\,\t_1
{}_\alpha y^\alpha+i\tau_2\, \t_2 {}_\alpha y^\alpha-i(\tau_1
\t_1+(1-\tau_2)\,\t_2)_\alpha\pp^{\alpha}
+i(\tau_2-\tau_1)\t_2 {}_\alpha\t_1^\alpha}\,,\nn\\
&\Phi^{[2,3]}=\t_1 \t_2
\int\dr^3_{\triangle}\tau\,e^{i(1-\tau_1)\,\t_2 {}_\alpha
y^\alpha+i\tau_2\, \t_1 {}_\alpha y^\alpha-i(\tau_1
\t_2+(1-\tau_2)\,\t_1)_\alpha \pp^\alpha +i(1-\tau_3)\,\t_2
{}_\alpha \t_1^\alpha}\,,\label{Cww}
\end{align}
where we also use the notation \eqref{contr}.  Vertices
$\Upsilon(\go, C, C)$ can be found from \eqref{wver} as easily
\be
\Upsilon(\go, C, C)=-\left(W_{1}(z'; y, \vec\y)*C-C*W_{1}(z'; -y,
\vec\y)\right)\Big|_{z'=-y}
\ee
with the final result being
\begin{align}
&\Phi^{[1]}=-\t y\int\dr^3_{\triangle}\tau\, e^{i(\tau_2\, \pp_2+
(1-\tau_2)\,\pp_1)_\alpha\t^\alpha+i((1-\tau_1)\, \pp_2+\tau_1\,\pp_1+(1-\tau_3)\,\t)_\alpha y^\alpha}\,,\\
&\Phi^{[2]}=\t y\int\dr^3_{\triangle}\tau\, e^{i(\tau_2\, \pp_2+
(1-\tau_2)\,\pp_1)_\alpha\t^\alpha+i((1-\tau_1)\, \pp_2+\tau_1\,\pp_1-(\tau_1-\tau_2)\,\t)_\alpha y^\alpha}+\nn\\
&\qquad+\t y\int\dr^3_{\triangle}\tau\, e^{i(\tau_2\, \pp_1+
(1-\tau_2)\,\pp_2)_\alpha\t^\alpha+i((1-\tau_1)\, \pp_1+\tau_1\,\pp_2+(\tau_1-\tau_2)\,\t)_\alpha y^\alpha}\,,\\
&\Phi^{[3]}=-\t y\int\dr^3_{\triangle}\tau\, e^{i(\tau_2\, \pp_1+
(1-\tau_2)\,\pp_2)_\alpha\t^\alpha+i((1-\tau_1)\,
\pp_1+\tau_1\,\pp_2-(1-\tau_3)\t)_\alpha y^\alpha}\,.
\end{align}
These simple examples above illustrate a few general phenomena
typical of all orders. Namely: (i) the exponential part of the
vertices never has contractions $\pp_i\cdot \pp_j$. Moreover,
there is no a single contraction $y\cdot \pp_i$ within
$\Phi^{[\delta_1,\delta_2]}$; and (ii) integration over $\tau$'s
goes along a simplex. At higher orders the integration domain
includes a peculiar structure of two hypersimplices. For example,
it is not difficult to extract the following any order vertex
explicitly
\be\label{verC}
\Phi^{[1]}_{n+1}=-(W^{(n)}_{\go C\dots\, C}\,(z';
Y)*C)_{z'=-y}=-W^{(n)}_{\go C\dots\, C}\,(-y; y-\pp_{n+1},
\vec\y)e^{-iy\pp_{n+1}}\,.
\ee
To do so we need $W$ to the $n^{th}$ order. It can be conveniently
found using that $W\in {\mathbf{C}}^{0}$ that makes representation
\eqref{f0} particularly useful. The concise form follows from the
iterative equation
\be
W^{(n)}=-\Delta(W^{(n-1)}*\Lambda)\,,
\ee
which gives us the final result (see Appendix B)
\be\label{Wn}
W^{(n)}=(z\t
)^n\int_{0}^{1}\dr\tau\,\tau^{n-1}(1-\tau)\int_{\mathcal{D}}
e^{-i(1-\tau)r_n\,y\t+i\tau\,z(y+B_n)+i\tau r_n\,B_n\t +ic_n}\,,
\ee
where
\begin{align}
&B_n=\sum_{i=1}^{n}\gl_i\pp_i+
\left(1+\sum_{i<j}^{n}(\gl_i\nu_j-\nu_i\gl_j)\right)\t\,,\label{Bn}\\
&c_n=\sum_{i=1}^{n} \nu_i\,\pp_i \t\,,\label{cn}\\
&r_n=1-\sum_{i=1}^{n}\nu_i\label{r}\,,
\end{align}
with the integration domain $\mathcal{D}$ being the product of two
constrained hypersimplices
\be\label{config}
\mathcal{D}=\triangle_n\times\triangle_n^*\,,
\ee
where
\be\label{D}
\triangle_n=\{0\leq \gl_i\leq 1;\quad \sum_{i=1}^{n}
\gl_i=1\}\,,\qquad \triangle_n^*=\{0\leq \nu_i\leq 1;\quad
\sum_{i=1}^n \nu_i\leq 1\}\,,
\ee
while the constraint has a shoelace form
\be\label{shoe}
\gl_{i+1}\nu_i-\gl_i\nu_{i+1}\geq 0\,,\qquad i=1\dots n-1
\ee

Note that the integration over $\gl_i$ goes along faces of
hypersimplex $\triangle_n$, while the integration over $\nu_i$
goes across its volume. Note also for $n=1$ the integration over
$\gl_1$ trivializes in a sense that $\gl_1\equiv 1$ as follows
from \eqref{D}. The appearance of integrals over hypersimplices
might not be too surprising. A suggestive analogy is the structure
of cubic coupling constants for helicities $h_i$ obtained in
\cite{Metsaev:1991mt}
\be
C_{h_1, h_2, h_3}=\ff{1}{\Gamma(h_1+h_2+h_3)}\,,
\ee
while the typical simplex integral features the gamma function in
the denominator as well
\be
\int_{\sum_i\nu_i\leq
1}\nu_1^{s_1-1}\dots\nu_n^{s_n-1}=\ff{\Gamma(s_1)\dots\Gamma(s_n)}{\Gamma(1+s_1+\dots
+s_n)}\,.
\ee

Another comment is since the result \eqref{Wn} belongs to class
$\mathbf{C}^0$, \eqref{f0}, it can be conveniently rewritten in
the factorized form with respect to operation \eqref{half}
\be
W^{(n)}=\int_{0}^{1}\dr\tau\,\tau^{n-1}(1-\tau)\int_{\mathcal{D}}
e^{-ir_n\,y\t+ic_n}\circledast (\t z)^n\,e^{i\tau\,z(y+B_n)}\,.
\ee
Vertex $\Phi_{n+1}$ is reproduced via \eqref{verC}. It carries the
structure of the hypersimplex integration domain just found.
Simplicity of the final result is noteworthy. All vertices
including those from \eqref{ups1} have similar form based on the
unique configuration space \eqref{config}. For example, from
\eqref{wwver} one obtains
\be
\Phi^{[1,2]}_n=-W^{(n)}(-\t_1, y+\t_1, \vec\y)\,e^{-i y\t_1}
\ee
with $W^{(n)}(z, y, \vec\y; \t_2, \pp_i)$ from \eqref{Wn}. It
would be interesting to elaborate more on the geometric properties
of the obtained integrals.

The resulting vertices feature a particular simple transformation
under shifts $\pp_i\to \pp_i+a$, where $a$ is an arbitrary spinor.
Specifically, one observes using \eqref{r} and \eqref{D} that
\begin{align}
&B_{n}(\pp_i+a, \t)=B_{n}(\pp_i, \t)+a\,,\label{shB}\\
&c_{n}(\pp_i+a, \t)=c_{n}(\pp_i+a, \t)+(1-r_n)\,a\t\,.\label{shc}
\end{align}
These relations underlie the so-called shift symmetry of the
interaction vertices, which we discuss in the next section.
Technically, most of the properties discussed above are literally
coincide with the analysis from \cite{Didenko:2022qga}, where
reader can find more details.

\subsection{Locality}

\paragraph{Spin-locality} The problem of locality of the
HS interactions in $d$ dimensions boils down to the analysis of
the number of Lorentz index contractions\footnote{Recall, the
rectangular Young diagram $C^{a(s), b(s)}$ corresponds to a spin
$s$ Weyl tensor, which is a physical field, while $C^{a(s+n),
b(s)}$ is its descendant made of the derivatives for $n>0$. The
number of these derivatives grows with $n$.} of various $C$'s
within vertices $\Upsilon(\go, C\dots C)$ and $\Upsilon(\go, \go,
C\dots C)$. This number can be arbitrarily large even for fixed
spins. Indeed, as the first row in \eqref{C} is unbounded, there
can be infinitely many contractions of the form, for example
\be
\sum_{m} g(m, s_1, s_2)\, C^{a(m)\dots,\,
b(s_1)}C_{a(m)\dots,}{}^{b(s_2)}\,.
\ee
Whenever this happens the vertex is said to be spin nonlocal. An
infinite length of the first row of \eqref{C} is supported by
infinitely many $y^{\al}\vec\y_{\al}$ contracted with it.
Therefore, nonlocality can be rephrased using the language of
$\Phi$ from \eqref{ups1} and \eqref{ups2} as a nonpolynomial
dependence on various contractions $\pp_i\cdot \pp_j$. So, the
vertex is said to be {\it spin-local} if $\Phi$ contains no
nonpolynomial  contractions $\pp\cdot \pp$ \cite{Gelfond:2018vmi},
\cite{Gelfond:2019tac}. Let us note, that dependence on $\t\cdot
\t$ and $\t\cdot \pp$ is irrelevant for spin-locality. This is due
to the fact that 1-forms $\go$ belong to a finite-dimensional
module of the HS algebra for fixed spin \eqref{w} as opposed to
$C$, which is infinite dimensional \eqref{C}.

\paragraph{Spin ultralocality} Another important concept
introduced in \cite{Didenko:2018fgx} is spin {\it ultralocality}.
Suppose $\Phi$ from \eqref{ups1} depends on contractions $y\cdot
\pp_i$. If such a dependence is at most polynomial, then the
corresponding vertex is called spin {\it ultralocal}. The meaning
of this concept is roughly the following. Imagine \eqref{ups1} is
spin-local but not ultralocal. This implies that the number of
various contractions is finite for fixed spins but grows with spin
sufficiently fast. As explained in \cite{Didenko:2022qga}, in
order for vertices \eqref{eqC} to be spin-local \eqref{ups1}
should be spin-ultralocal, which means that $\Phi^{[\gd_1,
\gd_2]}$ must contain no nonpolynomial $y\cdot \pp$ contractions.
These, if present, imply that while vertex is still spin local,
the depth of contractions of first rows between various $C$'s
grows with spins. This ruins spin-locality of \eqref{ups2} at the
next order in perturbation via integrability condition
\cite{Didenko:2022qga}. In addition, it was shown in
\cite{Gelfond:2019tac} in $d=4$ that if a vertex is
spin-ultralocal, then it is space-time spin-local in the usual
sense.

To summarize, (non)locality of the vertex constrained to a fixed
set of spins $s_i$ depends on function $\Phi$. To be
spin-nonlocal, $\Phi$ must contain at least one nonpolynomial
contractions $\pp_k\cdot \pp_l$. This is equivalent to having
infinitely many contractions with respect to first rows of two
$C$'s. Notice, that the presence of star product $\star$ in
\eqref{ups1} and \eqref{ups2} that acts on variable $\vec\y$
yields no such contractions and therefore can not affect locality.
The important property of equations \eqref{xeq}-\eqref{dCeq} is
that the off shell vertices they generate in \eqref{eqw} and
\eqref{eqC} are naturally spin-ultralocal and spin-local,
respectively. Therefore, the problem of the HS locality on shell
resolves into the algebraic one of a proper factorization of the
trace ideal. We do not pursue on shell (non)locality here leaving
this analysis for the future.

Spin-locality of traceful $\Upsilon$'s follows literally from the
analysis of \cite{Didenko:2022qga}. The only difference is one has
to replace $\bar{*}$ with $\star$ leaving the rest as is. For
details we refer the reader to section 5 in
ref.\cite{Didenko:2022qga}. Here we state the final result.
Namely, $\Phi^{[\gd_1, \gd_2]}$ contains neither $\pp\cdot y$, nor
$\pp\cdot \pp$ contractions which implies that $\Upsilon$'s from
\eqref{eqw} are ultralocal and therefore, according to
\cite{Gelfond:2019tac}, are space-time spin-local at any order.
Analogously, it follows that $\Phi^{[\gd]}$ carry no $\pp\cdot
\pp$ contractions while vertices from \eqref{eqC} are spin-local
off shell.

\subsection{Shift symmetry} Remarkable property of vertices
\eqref{ups1} and \eqref{ups2} is particularly simple
transformation laws under a shift of parameters $\pp_i$ and
oscillator $y$. Specifically, as was shown in
\cite{Didenko:2022qga} and as is clear from \eqref{shB},
\eqref{shc} the following transformations take place
\begin{align}
&\Phi^{[\gd_1, \gd_2]}(y-a; \t_1, \t_2,
\pp_i+a)=e^{i(\t_1+\t_2)^\alpha
a_\alpha}\Phi^{[\gd_1, \gd_2]}(y; \t_1, \t_2, \pp_i)\,,\label{sh2}\\
& \Phi^{[\gd]}(y; \t, \pp_i+b)=e^{i(\t+y)^\alpha
b_\alpha}\Phi^{[\gd]}(y; \t, \pp_i)\,,\label{sh1}
\end{align}
where $a$ and $b$ are arbitrary spinor parameters. In particular,
the above shift transformations leave vertices unaffected for
$a_{\al}=\mu (\t_1+\t_2)_{\al}$ and $b_{\al}=\nu (\t+y)_{\al}$
with arbitrary numbers $\mu$ and $\nu$. The proof is based on the
transformations \eqref{shB} and \eqref{shc} in the perturbative
expansion. A more detailed derivation can be found in
\cite{Didenko:2022qga}.

The symmetry transformations \eqref{sh2}, \eqref{sh1} have been
already observed in four dimensions and dubbed {\it shift
symmetry}. Hence, shift symmetry has a straightforward
generalization to any $d$. This observation is closely related to
the earlier analysis of \cite{Gelfond:2018vmi}, where it was shown
how certain parameter shifts of the so-called shifted homotopies
reduce the degree of nonlocality of HS vertices. While we use no
shifted homotopies in our studies, the idea to look at what shifts
in \eqref{sh2}, \eqref{sh1} might lead up to was to a large extend
inspired by the result \cite{Gelfond:2018vmi}. Since \eqref{ups1}
and \eqref{ups2} have meaning of the Fourier transformed HS
vertices, the symmetry naturally acts in the Fourier space.
However, its physical interpretation remains unclear. The recent
analysis of \cite{Didenko:2022eso} has revealed that spin-locality
is intimately related to shift symmetry. Given that, it would be
interesting to see whether or not shift symmetry in $d$ dimensions
goes through the on shell factorization.

\section{On shell reduction}\label{on-shell}

Equations \eqref{xeq}-\eqref{dCeq} with \eqref{lambda} are shown
to be consistent. They generate unfolded equations \eqref{eqw} and
\eqref{eqC}. As has been mentioned, the latter describe HS field
dynamics not until a proper factorization condition over the
tracefull ideal is imposed. Recall, that at the free level the
ideal is generated by the Howe dual $sp(2)$, \eqref{id}, provided
the HS fields $\go$ and $C$ are $sp(2)$ singlets
\be\label{Cyng}
[t_{\al\gb}^0, C]_*=[t_{\al\gb}^0, \go]_*=0\,,
\ee
Equation \eqref{Cyng} guarantees the spectrum to consist of the
two-row Young diagrams \eqref{w} and \eqref{C}, which, however,
are not traceless yet. To make them traceless one strips off the
ideal \eqref{id}. This simply implies a subtraction of traceful
components of Young diagrams in practice.

Now, nonlinear equations \eqref{eqw} and \eqref{eqC} that we can
denote $A_1=0$ for 1-forms and $A_0=0$ for 0-forms,
correspondingly, span two-row Young diagrams as well and,
therefore, one should have
\be\label{eqsingl}
[A_1, t^0_{\al\gb}]_*=[A_0, t^0_{\al\gb}]_*=0\,.
\ee
A seemingly natural idea is to set equations $A_{0,1}$ on-shell by
assuming fields $\go$ and $C$ properly traceless and then by
dropping terms associated with traceful contributions of the form
$t^{0}_{\al\gb}*A^{\al\gb}=A^{\al\gb}*t^{0}_{\al\gb}$, where
$A^{\al\gb}$ are chosen in such a way, so as to make e.g., eq.
\eqref{eqw} totally traceless. The procedure just prescribed
however does not lead to consistent interaction, because in adding
terms from the ideal one has to make sure that the resulting
equations remain consistent up to terms from the same ideal. For
the ideal in question this is not granted in interactions since
the off shell HS algebra along with its ideal get deformed. In
other words, the corresponding Howe dual $sp(2)$ generators
\eqref{singlet} receive field-dependent corrections
\be\label{tper}
t_{\al\gb}=t^0_{\al\gb}+O(C)\,.
\ee
So, the problem that one would like to address is whether there
still exists an algebra spanned by (possibly) field-dependent
generators that satisfy commutation relations \eqref{sp20}
\be\label{sp2}
[t_{\al\gb},
t_{\gga\gd}]_*=\gep_{\al\gga}t_{\gb\gd}+\gep_{\gb\gga}t_{\al\gd}+
\gep_{\al\gd}t_{\gb\gga}+ \gep_{\gb\gd}t_{\al\gga}\,,
\ee
and that allows for a consistent truncation of the associated
trace ideal at the level of the nonlinear equations \eqref{eqw}
and \eqref{eqC}.

\subsection{Global $sp(2)$ and the quotienting}

The described problem can be solved systematically if it is noted
that the singlet condition \eqref{singlet} can be viewed as a
requirement of existence of the global symmetry $sp(2)$ at the
level of the off shell equations. While deformed, it should still
be there in interactions to guarantee proper degrees of freedom.
Once generating system \eqref{xeq}-\eqref{dCeq} reproduces
\eqref{eqw}-\eqref{eqC}, we require the $sp(2)$ to be its global
symmetry,
\be
\gd_t W=\gd_t\Lambda=\gd_t C=0\,.\label{spglob}
\ee
Using explicit \eqref{locW}-\eqref{locC} one finds the
corresponding generators $t_{\al\gb}$ satisfy
\begin{align}
&\dr_x t_{\al\gb}+[W, t_{\al\gb}]_*=0\,,\label{22}\\
&\dr_z t_{\al\gb}+[\Lambda, t_{\al\gb}]_*=0\,,\label{33}\\
&\left(t_{\al\gb}(z'; y, \vec\y)*C-C*t_{\al\gb}(z'; -y,
\vec\y)\right)\Big|_{z'=-y}=0\,,\label{tcons}
\end{align}
where one should also remember \eqref{sp2}. Recall, \eqref{tcons}
is a consequence of \eqref{33}. Notice, that neither $W$ nor
$\Lambda$ are $sp(2)$ singlets in a sense of \eqref{singlet}.
However, the equations are. This can be shown using \eqref{22} and
\eqref{33} provided $t_{\al\gb}\in \mathbf{C}^0$
\begin{align}
&[\dr_x W+W*W, t_{\al\gb}]_*=0\,,\label{Wt}\\
&[\dr_z W+\{W,\Lambda\}_*+\dr_x\Lambda,
t_{\al\gb}]_*=0\,,\label{Lt}
\end{align}
where $[A,B]_*:=A*B-B*A$.
Following \cite{Vasiliev:2003ev} one wishes to drop terms from the
ideal at the level of the generating equations to set it on shell.
This is equivalent to adding an arbitrary ideal contribution to
the generating system \eqref{xeq}-\eqref{dCeq} so, that its
consistency gets ruined in its ideal part only. If this is the
case, then the factorization results in a consistent on shell
system from the factor space. To be a bit more specific, consider
eq. \eqref{eqw}, for example, that we have denoted by $A_1$
already. Being an $sp(2)$ singlet, thanks to \eqref{Wt}, it is an
element of the large HS off shell algebra
\be
[A_1, t_{\al\gb}]_*=0\,.
\ee
Decomposing
\be\label{Wid}
A_1=\mathbf{A}+A^{\mathbf{id}}\,,\qquad
A^{\mathbf{id}}=A^{\al\gb}*t_{\al\gb}=t_{\al\gb}*A^{\al\gb}\,,
\ee
where $\mathbf{A}$ is any particular representative of the HS
algebra (factor algebra), while $A^{\mathbf{id}}$ is an element of
the ideal. Both should commute with $t_{\al\gb}$
\be
[\mathbf{A}, t_{\al\gb}]_*=[A^{\mathbf{id}}, t_{\al\gb}]_*=0\,.
\ee
In that case the equation of motion $A_1=0$ entails
\be
\mathbf{A}=0\,,
\ee
which is equivalent to dropping off the part associated with the
ideal from equations. In practice, once the $z$-dependence of $W$
is resolved via \eqref{zeq}, eq. \eqref{eqw} becomes
$z$-independent. The natural representative for physical field
$\go$ is $z$-independent too. Hence, one has to require the ideal
part to be $z$-independent
\be
\dr_z(A^{\al\gb}*t_{\al\gb})=0\,.
\ee
Let us check now if the addition of $A^{\mathbf{id}}$ of the form
\eqref{Wid} to \eqref{eqw} is indeed consistent. To this end
consider
\be\label{weqid}
\dr_x W+W*W+A^{\al\gb}*t_{\al\gb}\simeq 0\,,
\ee
where $\simeq$ means equality up to terms that belong to the
ideal. Hitting with $\dr_x$ on \eqref{weqid} gives
\be
t_{\al\gb}*\mathrm{D}_x A^{\al\gb}\simeq 0\,,\qquad
\mathrm{D}_x=\dr_x+[ W,\bullet]\,,
\ee
which do belong to the ideal, since $\mathrm{D}_x t_{\al\gb}=0$
and $t_{\al\gb}*A^{\al\gb}=A^{\al\gb}*t_{\al\gb}$ and, therefore,
$t_{\al\gb}*\mathrm{D}_x A^{\al\gb}=\mathrm{D}_x
A^{\al\gb}*t_{\al\gb}$ is a two-sided ideal.

A similar analysis should be carried out for the rest of equations
\eqref{xeq}-\eqref{dCeq}. We do not do it here as we plan to
consider it in detail elsewhere.

\subsection{Explicit form of $sp(2)$ generators}
Let us collect all the conditions for $t_{\al\gb}$ together
\begin{align}
&[t_{\al\gb},
t_{\gga\gd}]_*=\gep_{\al\gga}t_{\gb\gd}+\gep_{\gb\gga}t_{\al\gd}+
\gep_{\al\gd}t_{\gb\gga}+ \gep_{\gb\gd}t_{\al\gga}\,,\label{1}\\
&\dr_x t_{\al\gb}+[W, t_{\al\gb}]_*=0\,,\label{2}\\
&\dr_z t_{\al\gb}+[\Lambda, t_{\al\gb}]_*=0\,.\label{3}
\end{align}
While the differential equations \eqref{2} and \eqref{3} are
consistent and must have solutions, {\it a priori} it is not
guaranteed that the $sp(2)$ condition \eqref{1} can be satisfied.
There is an elegant explanation on as to why the proper $sp(2)$
exists within the approach of \cite{Vasiliev:2003ev}. Its origin
is the algebra of the deformed oscillators that underlies
Vasiliev's master equations. It generates the required $sp(2)$ in
the covariant way in terms of the field analogous to $\Lambda$ via
its quadratic star-product combinations. Our approach is different
in some aspects. In particular, the star product operation
\eqref{limst} leads to an ill-defined product
$\mathbf{C}^1*\mathbf{C}^1$ in contrast with the Vasiliev case.
Since $\Lambda\in\mathbf{C}^1$, there is no hope we can define
quadratic combinations out of it. Nevertheless, if equations
\eqref{xeq}-\eqref{dCeq} describe a certain deformation of the
free off shell HS unfolded equations, the global symmetry of which
is $sp(2)$ spanned by \eqref{sp20}, then one should expect a
relevant deformation in interactions. Baring this option in mind
one can try to look at solutions of \eqref{1}-\eqref{3} in
perturbations in $C$ \eqref{tper}.

At order zero it is easy to see that $t^0$ of \eqref{singlet}
solves all the above conditions. Indeed, \eqref{1} is fulfilled by
its definition \eqref{sp20}. Equation \eqref{2} is satisfied
because $t^0$ is space-time constant, while $W^0=\go(Y)$ at this
order is an $sp(2)$ singlet, \eqref{Cyng}. Equation \eqref{3} is
satisfied since $t^0$ is $z$-independent, while the second term in
\eqref{3} is of the order $O(C)$.

\paragraph{First order} To proceed to the next order, let us  resolve the undeformed
singlet condition \eqref{singlet} explicitly. To this end we use
convention \eqref{contr} and introduce the following shorthand
notation
\begin{align}
&\eta_{ab}\y_{\al}^{a}\y_{\gb}^{b}:=\vec\y_{\al}\cdot\vec\y_{\gb}\,,\label{notation}\\
&\y_{\al}^a\y^{\al\,b}:=\vec\y\otimes\vec\y\,.
\end{align}
It is straightforward to check using \eqref{spop}, that functions
of the form
\be\label{CC}
C(y, \vec\y|\,x):=C(\vec\y y; \vec\y\otimes\vec\y|\,x)
\ee
are manifest $t^0$-singlets. At order $O(C)$ eq. \eqref{2} boils
down to
\be\label{t1st}
\dr_z t^{1}_{\al\gb}=[t^{0}_{\al\gb}, \Lambda]_*\,.
\ee
Its consistency \eqref{tcons} is just the free $sp(2)$-singlet
condition
\be
[t^0_{\al\gb}, C]_*=0\,,
\ee
which is solved by \eqref{CC}. Therefore, solution of \eqref{t1st}
exists and we can write it down using the standard conventional
homotopy. Substituting \eqref{lambda} into \eqref{t1st} and using
\be\label{tact}
[t^{0}_{\al\gb}, \bullet]_*=\vec\y_{\al}\cdot\ff{\p}{\p
\vec\y^{\gb}}+y_{\al}\ff{\p}{\p y^{\gb}}-i\ff{\p}{\p
y^{\al}}\ff{\p}{\p z^{\gb}}+(\al\leftrightarrow\gb)
\ee
gives (upon changing the integration variable)
\be\label{t1res}
t^1_{\al\gb}=-z_{\al}z_{\gb}\int_{0}^{1}\dr\tau\,\tau(1-\tau)\,C(-\tau\vec\y
z; \vec\y\otimes\vec\y)\,e^{i\tau zy}\,.
\ee
A possible freedom in $z$-independent function of this solution is
fixed to zero by \eqref{1}. Indeed, suppose $t^{1}_{\al\gb}$ ruins
$sp(2)$ commutation relations \eqref{1} up to $O(C)$. Using that
antisymmetric pair of spinor indices is proportional to the
$sp(2)$ epsilon, we have at this order
\be\label{chk1}
[t_{\al\al}, t_{\gb\gb}]_*\Big|_{O(C)}=[t^0_{\al\al},
t^1_{\gb\gb}]_*+[t^1_{\al\al},
t^0_{\gb\gb}]_*=2\gep_{\al\gb}S_{\al\gb}\,,
\ee
where $S_{\al\gb}=S_{\gb\al}$ and we suppose $S_{\al\gb}\neq
t^1_{\al\gb}$. From \eqref{3} one concludes, however, that
\be
\dr_z S_{\al\gb}=[t^{0}_{\al\gb}, \Lambda]_*\,.
\ee
Therefore, $S_{\al\gb}=t^1_{\al\gb}+\gs_{\al\gb}(y, \vec\y)$,
where $\gs$ is some arbitrary function. Setting $z=0$ in
\eqref{chk1} one finds that its left hand-side vanishes, while the
$\gs$-contribution on the right-hand side survives; thus,
$\gs_{\al\gb}=0$. Notice also that $t^{1}\in {\mathbf{C}}^{0}$ as
required.

\paragraph{Second order} Let us now inspect \eqref{3} at order $O(C^2)$
\be\label{t2eq}
\dr_z t^2_{\al\gb}=[t^1_{\al\gb}, \Lambda]_*\,.
\ee
Consistency $\dr_z^2=0$ by virtue of \eqref{iden} constrains
\be\label{cons2}
\left(t^{1}_{\al\gb}(z', y)*C-C*t^{1}_{\al\gb}(z',
-y)\right)\Big|_{z=-y}=0\,.
\ee
We can check whether \eqref{cons2} is fulfilled by direct
computation that gives
\begin{align}
&\left(t^{1}_{\al\gb}(z,
y)*C\right)\Big|_{z=-y}=-y_{\al}y_{\gb}\int_{0}^{1}\dr\tau\,\tau(1-\tau)\,
C(\tau\vec\y y; \vec\y\otimes\vec\y)\star C((1-\tau)\vec\y y; \vec\y\otimes\vec\y )\,,\label{tC}\\
&\left(C*t^{1}_{\al\gb}(z,
-y)\right)\Big|_{z=-y}=-y_{\al}y_{\gb}\int_{0}^{1}\dr\tau\,\tau(1-\tau)\,
C((1-\tau)\vec\y y; \vec\y\otimes\vec\y)\star C(\tau\vec\y y;
\vec\y\otimes\vec\y )\,,\label{Ct}
\end{align}
where $\star$ is a part of the Moyal product \eqref{moyal}
associated with $\vec\y$ variables \eqref{starvec}. Note that the
prefactor in \eqref{tC} and \eqref{Ct} is symmetric under $\tau\to
1-\tau$ and therefore the two terms above are the same as
consistency \eqref{cons2} holds. Less expected is that the whole
right-hand side of \eqref{t2eq} vanishes for $t^1$ from
\eqref{t1res}
\be
[t^{1}_{\al\gb}, \Lambda]_*=0
\ee
and, therefore, $t^{2}$ is $z$-independent. The cancellation in
\eqref{t2eq} can be observed by straightforward star-product
calculation using the singlet ansatz \eqref{CC} along with
symmetry $\tau\to 1-\tau$ in the measure of \eqref{t1res}.
Moreover, similar analysis results in (see Appendix C)
\be
[t^{1}_{\al\gb}, t^{1}_{\gga\gd}]_*=0\,.
\ee
This suggests higher order in $C$ corrections to $t_{\al\gb}$ are
absent as the final result that satisfies both \eqref{1} and
\eqref{3} is at most linear in $C$
\be\label{t}
\boxed{t_{\al\gb}=\ff{1}{4i}(\vec\y_{\al}\cdot
\vec\y_{\gb}+y_{\al}y_{\gb})-z_{\al}z_{\gb}\int_{0}^{1}\dr\tau\,\tau(1-\tau)\,C(-\tau\vec\y
z; \vec\y\otimes\vec\y)\,e^{i\tau zy}\,.}
\ee
There are no higher order in $C$ corrections as \eqref{t} is
all-order exact.

\paragraph{Covariant constancy}
Solution \eqref{t} is shown to satisfy all the required conditions
except for \eqref{2}. Note, there is no freedom left in \eqref{t}.
Given the freedom in solutions of \eqref{zeq} for $W$, eq.
\eqref{2} can be satisfied for some particular choice of $W$. We
now prove that $t_{\al\gb}$ from \eqref{t} satisfies the remaining
covariant constancy condition \eqref{2} provided field $W$ is
fixed canonically \eqref{canembd}.

Suppose now eq. \eqref{2} is not valid and therefore,
\be\label{xR}
\dr_x t_{\al\gb}+[W,t_{\al\gb}]_*=R_{\al\gb}\neq 0\,.
\ee
Using \eqref{3} one finds the following consistency constraint for
$R$
\be\label{zR}
\dr_z R_{\al\gb}+[\Lambda, R_{\al\gb}]_*=0\,.
\ee
Let us analyze \eqref{zR} in perturbations. At zeroth order we
know that $R^{{0}}_{\al\gb}=0$ and, therefore, it follows from
\eqref{zR}
\be
R^{1}_{\al\gb}=R^{1}_{\al\gb}(y, \vec\y)
\ee
is $z$-independent. Hence, $R^{1}$ can be found from \eqref{xR} by
setting $z=0$
\be
R^{1}_{\al\gb}=\left([\go, t^1_{\al\gb}]_*+[W_1,
t^0_{\al\gb}]_*\right)\Big|_{z=0}\,.
\ee
Let us show that
\be\label{Rzero}
([\go, t^1_{\al\gb}]_*)_{z=0}=([t^0_{\al\gb},
W_1]_*)_{z=0}\quad\Rightarrow\quad R^{1}_{\al\gb}=0\,.
\ee
To this end we use \eqref{tact}
\be
[t^0_{\al\gb}, W_1]_*\Big|_{z=0}=-\ff{i}{2}\left(\ff{\p}{\p
y^{\al}}\ff{\p}{\p z^{\gb}}+\ff{\p}{\p y^{\gb}}\ff{\p}{\p
z^{\al}}\right)W_1\Big|_{z=0}\,,
\ee
where $y\ff{\p}{\p y}$ in $t^0$ does not contribute at $z=0$ since
$W_1(z=0; Y)=0$. Then, from the $z$-evolution equation \eqref{zeq}
we have
\be
\ff{\p}{\p
z^{\al}}W_1\Big|_{z=0}=(\go*\Lambda_{\al}-\Lambda_{\al}*\go)_{z=0}\,,
\ee
where one also takes into account that $\dr_x\Lambda=0$ at $z=0$.
Note, $W_1$ contains two possible orderings $\go\star C$ and
$C\star\go$. If \eqref{Rzero} is valid, it should be so for each
ordering separately. Let us check ordering $\go\star C$ first. It
is convenient to use the Taylor form of $\go$ \eqref{wexp}
\be
\go(y, \vec\y)=e^{-iy \t}\go(y', \vec\y)\Big|_{y'=0}\,.
\ee
Straightforward $*$ calculation yields
\be
\ff{\p}{\p
y^{\al}}(\go*\Lambda_{\gb})_{z=0}=-i\t_{\al}\t_{\gb}\int_{0}^{1}\dr\tau\,
\tau(1-\tau)e^{-i(1-\tau)y \t}\go(y', \vec\y)\star C(-\tau \t,
\vec\y)\Big|_{y'=0}\,.
\ee
At the same time,
\be
(\go*t^{1}_{\al\gb})_{z=0}=-\t_{\al}\t_{\gb}\int_{0}^{1}\dr\tau\,
\tau(1-\tau)e^{-i(1-\tau)y \t}\go(y', \vec\y)\star C(-\tau \t,
\vec\y)\Big|_{y'=0}
\ee
which precisely matches \eqref{Rzero} for ordering $\go\star C$.
Similarly, the required cancellation takes place for ordering
$C\star\go$.

At second order we find from \eqref{zR} that once
$R^1_{\al\gb}=0$, then $R^2_{\al\gb}$ is $z$-independent and
therefore, from \eqref{xR} the latter can be found as
\be\label{R2}
R^{2}_{\al\gb}=\left([W_1, t^1_{\al\gb}]_*+[W_2,
t^0_{\al\gb}]_*\right)_{z=0}
\ee
The canonical $W_2$ is of $O(z^2)$ and, hence, once $t^0$ contains
no more than first derivative over $z$, the second term on the
right-hand sise of \eqref{R2} vanishes. Furthermore, before
setting $z=0$ each term from commutator $[W_1, t^1_{\al\gb}]_*$
can be shown to be of $O(z)$ by direct calculation. So, the first
term on the right of \eqref{R2} vanishes at $z=0$ just as well.
One then concludes that $R^2_{\al\gb}=0$. The above consideration
extends analogously to higher orders due to $W_n$ is at least
$O(z^2)$ for $n\geq 2$, provided the canonical choice of field
$\go$ is made. This proves \eqref{2} and provides the final simple
form of all-order generators \eqref{t}.

\section{Outlook and discussion}\label{vse}

We have shown that the Vasiliev type generating equations of
\cite{Didenko:2022qga} can be framed to describe interacting
symmetric HS gauge fields in any dimension. Based on the off shell
HS algebra they provide a set of HS compatibility constraints. The
dynamical evolution originates from factorization over the trace
ideal governed by the Howe dual $sp(2)$ that reduces the system
down to its mass shell. Since the off shell HS algebra gets
deformed in interaction, the factorization procedure becomes
highly non-trivial and {\it a priori} unknown. To solve this
problem we explicitly find field-dependent generators of $sp(2)$
commuting with the generating equations. These allow one to set
equations on shell by stripping the traceful contributions off.
The remarkable feature of the generators obtained is a very simple
linear in field form \eqref{t}, which is one of the main results
of the current research. The cancellation of higher-order field
corrections was not obvious and came as a surprise. This is not
typical of the original Vasiliev case as field dependence of
$sp(2)$ is not bounded by linear terms in general. A concise form
of the generators found makes on shell analysis feasible, while
the HS locality issue in $d$ dimensions amenable.

The main objective of using generating equations of the proposed
type is all-order spin-locality of the (off shell) HS vertices
manifestly available via canonical choice of field variables
\eqref{canembd}. The respective vertices were shown to be
space-time spin-local in \eqref{eqw} and spin-local in
\eqref{eqC}. To arrive at this result one uses the standard
contracting homotopy resolution operator all the way in
perturbations without any further field redefinitions, thus
implementing the canonical embedding. The manifest form of off
shell HS vertices have been obtained at any order in a neat form
of integrals over hypersimplices, the dimension of which grows
with the order of perturbation theory.

While we have not carried out the on shell reduction in this
paper, an important remark in this regard is in order. The
canonical choice \eqref{canembd}, being irrelevant to consistency
of the generating equations plays an essential role in
factorization of the trace ideal. Namely, in manifestly deriving
$sp(2)$ generators \eqref{t}, it was the canonical embedding that
guaranteed the proposed ansatz to satisfy all necessary
constraints.

Let us comment now on the differences between generating equations
\eqref{xeq}-\eqref{dCeq} and the Vasiliev ones from
\cite{Vasiliev:2003ev}. The minor difference is we do not double
variables $Y$'s at nonlinear level, rather add a two-component
$z_{\al}$ which is sufficient to grasp at nonlinear level. We do
not take advantage of the AdS covariant setting of the original
equations \cite{Vasiliev:2003ev} either as we prefer to stay in
the Lorentz frame. The major departure is the choice of the
$z$-commuting large algebra \eqref{limst} which significantly
shifts the whole setting.\footnote{An observation of the limiting
star product that leads to \eqref{limst} in \cite{Didenko:2019xzz}
was stimulated by the results of \cite{DeFilippi:2019jqq}, where a
relation of specific homotopy shifts to star-product ordering was
pointed out (see also \cite{Iazeolla:2020jee} and
\cite{DeFilippi:2021xon} for further discussion on orderings).
Equation \eqref{limst} plays a pivotal role in the analysis of
\cite{Didenko:2022qga}.} As explained in detail in
\cite{Didenko:2022qga} this choice does not quite live up to the
original Vasiliev construction causing early infinities in
interactions. This is one of the reasons why Vasiliev's $Z$ do not
commute.\footnote{In \cite{Bekaert:2004qos} it was stressed that
$Z$'s should be noncommuting for a yet another reason.}
Nevertheless, it is precisely the star product \eqref{limst} that
effectively manifests while evaluating HS vertices
\cite{Didenko:2019xzz} constrained by locality using the original
Vasiliev framework. This apparent contradiction is resolved by
postulating in \cite{Didenko:2022qga} equations of the form
\eqref{xeq}-\eqref{dCeq}. As compared to the original Vasiliev
ones, they lack zero-form module $B$, while HS vertices of
\eqref{eqC} are generated by \eqref{dCeq}. Consistency rests on
the remarkable, yet so far poorly understood {\it projective
identity} \eqref{iden}. This identity is strongly tied to a
particular solution \eqref{lambda} of \eqref{dzlambda}. On the
contrary, there is no formal solution selection for the analog of
field $\Lambda$ within the Vasiliev equations.

Another important difference is the lack of quadratic contribution
$\Lambda*\Lambda$ in \eqref{zeq} in contrast with the similar
Vasiliev case, where not only it is present due to consistency, it
plays a fundamental role. This term is responsible for the local
Lorentz symmetry of the spinorial $d=4$ equations
\cite{Vasiliev:1992av}, that would be not otherwise guaranteed. It
as well plays a key role in the on shell projection of
\cite{Vasiliev:2003ev}. In both cases the deformed oscillator
algebra of the aforementioned quadratic contribution does the job.
This missing ingredient challenges the global $sp(2)$, which is
necessary to make sense of our equations on shell. Nevertheless,
the global $sp(2)$ does exist in our case, although we find its
presence not obvious beforehand. It is also worth recalling that
the corresponding generators are of remarkably simple form for the
canonical definition of the physical field \eqref{canembd}. The
latter choice then again is in consonance with spin-locality.

Despite the differences we believe equations
\eqref{xeq}-\eqref{dCeq} should follow from Vasiliev's ones
perhaps nontrivially. It would be interesting to understand the
link between the two systems. It is not unlikely that the former
results from the latter upon star-product contraction along the
lines of Section 6 from \cite{Didenko:2019xzz} and a proper
renormalization of the Vasiliev vacuum state. Constrained by
associativity, however, the limiting procedure makes its
implementation beyond order $C^2$ challenging.\footnote{At order
$C^2$ equivalence of the two systems can be established using a
peculiar identity (6.29) of \cite{Didenko:2019xzz}.} One reason to
expect equations \eqref{xeq}-\eqref{dCeq} may appear as a result
of a certain contraction is the structure of functional class
\eqref{class}, which is a subclass of the one from
\cite{Gelfond:2019tac} designed to reconcile locality. However, in
our case the system has no room for a class bigger than
\eqref{class} at least in perturbations that indicates its
reductive origin.

In conclusion let us point out some interesting problems for the
future investigation
\begin{itemize}
\item Computation of on shell HS vertices at orders $C^2$ and
$C^3$ and beyond along with examination of spin-locality. The
developed formalism seems quite suitable for that, as on the one
hand HS vertices are shown to be space-time local to any order in
$C$ at the off shell level; thus one is left with a careful
analysis of the factorization condition. On the other hand, the
factorization procedure looks encouraging in view of remarkably
simple $sp(2)$ generators \eqref{t}. At this stage it is not clear
whether one should proceed along the lines of the quasiprojector
technique \cite{Vasiliev:2004cm}, \cite{Sagnotti:2004pod},
\cite{Bekaert:2004qos}, \cite{Joung:2014qya} (see also
\cite{Alkalaev:2019xuv} for the somewhat degenerate lower
dimensional case) in attacking this problem or some new tools
should be developed. Indeed, the action of quasiprojectors might
not be necessarily compatible with locality.

\item An intriguing HS shift symmetry that manifests itself within
the spin-local setting has not received an adequate explanation so
far (a technical reason is the presence of hypersimplex in the
integration domain). In particular, it is interesting to check if
it stands the on shell factorization. Given the results of
\cite{Didenko:2022eso}, where spin-locality was shown to follow
from the shift symmetry assumption, it seems likely to define a
class of proper on shell field representatives from factor space.

\item An immediate application of eqs. \eqref{xeq}-\eqref{dCeq}
could be the $3d$ HS interactions of Prokushkin and Vasiliev
\cite{Prokushkin:1998bq}, which are on shell due to their
spinorial setup. It would be interesting to revisit locality issue
for this theory.
\end{itemize}

\section*{Acknowledgments}
We are grateful to Kostya Alkalaev for useful discussions. We also
acknowledge Mikhail Povarnin for his careful reading of the
manuscript and pointing out a few typos in formulas in Section 3.
VE acknowledges the financial support from the Foundation for the
Advancement of Theoretical Physics and Mathematics ``BASIS''.

\newcounter{appendix}
\setcounter{appendix}{1}
\renewcommand{\theequation}{\Alph{appendix}.\arabic{equation}}
\addtocounter{section}{1} \setcounter{equation}{0}
 \renewcommand{\thesection}{\Alph{appendix}.}

\addcontentsline{toc}{section}{\,\,\,\,\,\,\,A. Classes of
functions}

\section*{A. Classes of functions}\label{AppA}

The important step is to identify a class of functions
${\mathbf{C}}^{r}$ that lives on all operations entering
\eqref{xeq}-\eqref{dCeq}. As the degree of $\dr z$- is a grading,
the class is graded correspondingly. We define $r=0, 1, 2$ to be a
label prescribed to zero-, one-, and two-forms, respectively.
Accordingly, we require
\begin{align}
&{\mathbf{C}}^{r_1}*{\mathbf{C}}^{r_2}\in
{\mathbf{C}}^{r_1+r_2}\,,\label{cl1}\\
&\dr_z {\mathbf{C}}^{r}={\mathbf{C}}^{r+1}\,.\label{cl2}
\end{align}
Conditions \eqref{cl1} and \eqref{cl2} have been analyzed in
\cite{Didenko:2022qga} with the following explicit result:
${\mathbf{C}}^{r}$ consists of the following functions
\begin{align}\label{class}
\phi(z, y; \vec\y,
\dr z)&=\\
&=\int_{0}^{1}\dr\tau\ff{1-\tau}{\tau}\int \ff{\dr u\,\dr
v}{(2\pi)^2}\, f\Big(\tau z+v, (1-\tau)(y+u); \vec\y,
\ff{\tau}{1-\tau}\dr z\Big)e^{i\tau
z_{\al}y^{\al}+iu_{\al}v^{\al}}\,,\nn
\end{align}
where $f(z,y; \vec\y, \dr z)$ is such that the integration is well
defined and is otherwise arbitrary analytic function. A convenient
way to visualize the above class is to use the source
parametrization which gives $\phi$ in different $\dr z$-sectors
\begin{align}
&1:\qquad\int_{0}^{1}\dr\tau\,\ff{1-\tau}{\tau}e^{i\tau
z_{\al}y^{\al}+i(1-\tau)A^{\al}y_{\al}+i\tau B^{\al}z_{\al}-i\tau
A^{\al}B_{\al}}\,,\label{f0}\\
&\dr z^\alpha:\qquad\int_{0}^{1}\dr\tau\, e^{i\tau
z_{\al}y^{\al}+i(1-\tau)A^{\al}y_{\al}+i\tau
B^{\al}z_{\al}+i(1-\tau) A^{\al}B_{\al}}\,,\label{f1}\\
&\dr z^\alpha \dr z_\alpha:\qquad\int_{0}^{1}\dr\tau\,\ff{\tau}{1-\tau}e^{i\tau
z_{\al}y^{\al}+i(1-\tau)A^{\al}y_{\al}+i\tau
B^{\al}z_{\al}+i(1-\tau) A^{\al}B_{\al}}\,.\label{f2}
\end{align}
Differentiating with respect to sources $A$ and $B$ and then
setting them to zero one reproduces various elements of
${\mathbf{C}}^{r}$. Freedom in coefficients contains arbitrary
functions of $\vec\y$. Note, that in order $\tau$-integration to
be well-defined, \eqref{f0} and \eqref{f2} should be at least
linear in $A$.

The important properties of functions from ${\mathbf{C}}^{r}$ are
as follows:
\begin{itemize}
\item While not immediately obvious, it has been shown in
\cite{Didenko:2022qga} that functions \eqref{class} exhaust all
possible perturbative solutions of \eqref{xeq}-\eqref{dCeq}. In
particular, $z$-independent zero-forms do belong to
${\mathbf{C}}^{0}$. Therefore, field redefinitions $\go\to
\go+F(\go, C\dots C)$ are not constrained by \eqref{class}. It is
also important that $\Delta (\Lambda*{\mathbf{C}}^{0})\in
{\mathbf{C}}^{0}$ and $\Delta ({\mathbf{C}}^{0}*\Lambda)\in
{\mathbf{C}}^{0}$, where $\Delta$ is the homotopy operator given
in \eqref{hom}.

\item There is an invariance (under proper rescaling) of functions
from ${\mathbf{C}}^{r}$ with respect to the following star-product
reordering operator
\be\label{ord}
O_{\gb}\phi(z,y; \vec\y)=\int\ff{\dr^2 u\dr^2 v}{(2\pi)^2}
\phi(z+v,y+\gb u; \vec\y)\exp (iu_{\al}v^{\al})\,.
\ee
It can be shown that
\be
O_{\gb}\phi(z,y; \vec\y)=\phi\left(\ff{z}{1-\gb}, y;
\vec\y\right)\,,\qquad\forall \phi\in {\mathbf{C}}^{r}\,,
\ee
where $\gb$ is an arbitrary number. It plays a role of a
reordering parameter of the original Vasiliev star product
\cite{Didenko:2019xzz}. The limiting star product \eqref{limst}
emerges from the Vasiliev one in the limit $\gb\to-\infty$. Eq.
\eqref{ord} says that \eqref{class} is a fixed point of the
re-ordering operator (upon a proper rescaling).

\item Lastly, product ${\mathbf{C}}^{r_1}*{\mathbf{C}}^{r_2}$ is
well-defined provided $r_1+r_2<2$ or one of the functions (from
${\mathbf{C}}^{0}$) is $z$-independent, while another one belongs
to ${\mathbf{C}}^{2}$.
\end{itemize}
There is a useful factorization formula for \eqref{f0}-\eqref{f2}
convenient in practice (\cite{Didenko:2022qga},
\cite{Didenko:2022eso}). Namely,
\be
e^{i\tau z(y+B)+i(1-\tau)yA-i\tau BA}=e^{iy A}\circledast e^{i\tau
z(y+B)}\,,
\ee
where
\be\label{half}
f(y)\circledast g(z,y)= \int f(y+u)g(z-v,y)\, e^{iuv}\,.
\ee
Note that $f(y)$ is $z$-independent.

\renewcommand{\theequation}{\Alph{appendix}.\arabic{equation}}
\addtocounter{appendix}{1} \setcounter{equation}{0}
\addtocounter{section}{1}
\addcontentsline{toc}{section}{\,\,\,\,\, B. Higher orders
off-shell}

\section*{B. Higher orders off shell}\label{AppB}

Here we derive eq. \eqref{Wn}. To do so we use the following
generating formula. For
\be\label{appgenf}
f=\int_{0}^{1}\dr\tau\ff{1-\tau}{\tau} e^{iy A}\circledast
e^{i\tau z(y+B)}
\ee
one has
\be
f*\Lambda=\dr
z^{\al}\int_{[0,1]^2}\dr\tau\,\dr\gs\,\ff{\gs}{1-\gs}\,e^{iyA}\circledast
\tau z_{\al}e^{i\tau z(y+\gs (\pp-A)+(1-\gs)B)}\,,\label{appfL}
\ee
where $\Lambda$ is given in \eqref{Lexp}, while \eqref{appgenf}
should be understood as generating in terms of sources $A$ and
$B$. In order to be well-defined there must be no contribution at
$B=0$. In other words, $f$ should be at least linear in $B$.

As we consider mostly left ordering $\go C\dots\,C$, we have at
order $n+1$
\be\label{Wnext}
W_{n+1}=-\Delta(W_n*\Lambda)\,,
\ee
where $\Delta$ is the standard contracting homotopy defined in
\eqref{hom} that suggests the following ansatz
\be
W_{n}=\int \mu_n\, e^{-ir_n\,y\t+ic_n}\circledast (z\t
)^n\,\tau^{n-1}(1-\tau)\,e^{i\tau\,z(y+B_n)}\,.
\ee
Using \eqref{appfL} and the identity
\be\label{int}
\int_{[0,1]^2}\dr\tau_1\dr\tau_2\, f(\tau_1\tau_2,
\tau_1)=\int_{[0,1]^2}\dr\tau\dr\rho\,\ff{1-\tau}{1-(1-\tau)\rho}f(\tau,
1-(1-\tau)\rho)
\ee
one finds from \eqref{Wnext}
\be
W_{n+1}=\int \mu_{n+1}\, e^{-ir_{n+1}\,y\t+ic_{n+1}}\circledast
(z\t)^{n+1}\,\tau^{n}(1-\tau)\,e^{i\tau\,z(y+B_{n+1})}\,,
\ee
where
\begin{align}
&\mu_{n+1}=\mu_n\,r_n\,\gs_{n+1}(1-\gs_{n+1})^{n-1}\,,\label{cond1}\\
&B_{n+1}=\gs_{n+1}(\pp_{n+1}+r_n\,\t)+(1-\gs_{n+1})B_n\,,\label{cond2}\\
&c_{n+1}=c_n+r_n(1-\rho_{n+1})(B_{n+1}\,\t)\,,\label{cond3}\\
&r_{n+1}=\rho_{n+1}r_n\,,\label{cond4}
\end{align}
where by $\gs_{n+1}$ and $\rho_{n+1}$ we denote integration
variables $\gs$ and $\rho$ from \eqref{int} that arise at the
stage $n+1$. The initial values are
\begin{align}\label{inval}
\mu_1=1\,,\quad \gs_1=1\,,\quad r_1=\rho_1\,,\quad
c_{1}=(1-\rho_1)(\pp_1\,\t)\,,\quad B_1=\pp_1+\t\,.
\end{align}
The integration goes along $\gs_i$ ($i\geq 2$), $\rho_i$ ($i\geq
1$) and $\tau$ -- all ranging $[0,1]$. The solution of
\eqref{cond2}-\eqref{cond4} is given by \eqref{Bn} and \eqref{cn}
that can be proven by induction. Indeed, assuming \eqref{Bn} and
\eqref{cn} hold at order $n$ and plugging these expressions
explicitly in \eqref{cond2}, \eqref{cond3} we have
\begin{align}
&B_{n+1}=\gs_{n+1}(\pp_{n+1}+\rho_1\dots\rho_n\,\t)+(1-\gs_{n+1})
\left(\sum_{i=1}^n \gl_i
\pp_i+\Big(1+\sum_{i<j}^n(\gl_i\nu_j-\nu_i\gl_j)\Big)\t\right)\,,\\
&c_{n+1}=\sum_{i=1}^n \nu_i\,(\pp_i\,\t)+\rho_1\dots\rho_n(1-\rho_{n+1})
\left(\gs_{n+1} \pp_{n+1}+(1-\gs_{n+1})\sum_{i=1}^n
\gl_i\pp_i\right)\t\,,
\end{align}
where
\be
1-\rho_1\dots\rho_n=\sum_{i=1}^n \nu_i\,,\; \;\;\sum_{i=1}^n \lambda_i=1\,
\ee
holds by inductive assumption. Introducing new variables
\begin{align}
&\gl_{n+1}'=\gs_{n+1}\,,\qquad
\gl_{i}'=(1-\gs_{n+1})\gl_i\quad\Rightarrow\quad \sum_{i'}
\gl_{i'}'=1\,,\label{gl}\\
&\nu_{i}'=\nu_i+\rho_1\dots\rho_n(1-\rho_{n+1})\gl_{i}'\,,\qquad
\nu_{n+1}'=\rho_1\dots\rho_n(1-\rho_{n+1})\gl_{n+1}'\label{nu}
\end{align}
\be\label{vcond}
1-\rho_1\dots\rho_{n+1}=\sum_{i'} \nu_{i'}'\,,\qquad
1-\rho_1\dots\rho_{n}=\sum_{i} \nu_{i}\,,
\ee
where index $i'$ ranges $n+1$ values and observing that
\be\label{primeid}
\gl_{i}'\nu_{j}-\nu_{i}\gl_{j}'=\gl_{i}'\nu_{j}'-\nu_{i}'\gl_{j}'
\ee
one easily finds
\begin{align}
&B_{n+1}=\sum_{i'}\gl_{i'}'
\pp_{i'}+\left(1+\sum_{i'<j'}(\gl_{i'}'\nu_{j'}'-\nu_{i'}'\gl_{j'}')\right)\t\,,\\
&c_{n+1}=\sum_{i'} \nu_{i'}'\,(\pp_{i'}\, \t)\,,\\
&r_{n+1}=1-\sum_{i'} \nu_{i'}'\,,
\end{align}
thus proving \eqref{Bn}, \eqref{cn} and \eqref{r}.

\paragraph{Jacobian} Let us show now that $(\gl,\nu)$ variables imply $\mu_{n+1}\times J=1$, where $J$ is the
Jacobian of $(\gs, \rho)\to (\gl, \nu)$ variable change. For $n=1$
one notes that $\mu_1=1$. For $n\geq 1$ we can show $\mu_n=1$. The
proof goes inductively. Assuming $\mu_n=1$, we then have from
\eqref{cond1}
\be\label{mu}
\mu_{n+1}=\rho_1\dots\rho_n\,\gs_{n+1}(1-\gs_{n+1})^{n-1}\,.
\ee
Variable change \eqref{gl}, \eqref{nu} yields its further
multiplication by the Jacobian arising from integration over the
following $\gd$-functions
\begin{align}
&J=\gd(\gl'_{i}-(1-\gs_{n+1})\gl_i)\gd(\nu_{n+1}'-\rho_1\dots\rho_n(1-\rho_{n+1})\gs_{n+1})\\
&\gd(\nu_{i}'-\nu_i-\rho_1\dots\rho_n(1-\rho_{n+1})\gl_{i}')\gd(1-\sum_i
\gl_i)\gd(1-\rho_1\dots\rho_n-\sum_{i}\nu_i)\,,\nn
\end{align}
with respect to $\gl_i$ and $\nu_i$. By integrating out
delta-functions we see that they exactly cancel out \eqref{mu}
\be
\mu_{n+1}\times J=1\,.
\ee

\paragraph{Integration domain} The configuration space is easily identified by induction. Assuming it is
given by \eqref{D}, \eqref{shoe} at order $n$, being the case at
$n=1$ (somewhat trivially, since $\gl_1=1$) and $n=2$ (regularly),
one sees from \eqref{gl} that $\gl$-hypersimplex extends to the
order $n+1$, while from \eqref{nu} and \eqref{vcond} the
$\nu$-hypersimplex holds at $n+1$ too. Now, let us show that the
shoelace constraint \eqref{shoe} holds at the next order. To see
this we first note, that \eqref{shoe} is equivalent to
\be\label{idshoe}
\gl_i\nu_j-\gl_j\nu_{i}\geq 0\,,\qquad i>j\,.
\ee
From \eqref{gl}, \eqref{nu}, \eqref{vcond} and \eqref{primeid} it
follows that if \eqref{idshoe} is true, then
\be
\gl_{i'}'\nu_{j'}'-\nu_{i'}'\gl_{j'}'\geq 0\,,\qquad i'>j'
\ee
and, therefore, \eqref{shoe} remains valid at order $n+1$.

\renewcommand{\theequation}{\Alph{appendix}.\arabic{equation}}
\addtocounter{appendix}{1} \setcounter{equation}{0}
\addtocounter{section}{1}
\addcontentsline{toc}{section}{\,\,\,\,\, C. $sp(2)$ proof}

\section*{C. $sp(2)$ proof}\label{AppC}

Let us check that \eqref{t} indeed generates $sp(2)$ commutation
relations \eqref{1}. We have
\be
t_{\al\gb}=t^0_{\al\gb}+t^{1}_{\al\gb}\,,
\ee
where $t^0$ is given in \eqref{singlet}, while $t^1$ is defined in
\eqref{t1res}. In order to see that \eqref{1} is fulfilled we are
left to check that
\be\label{prove}
[t_{\al\gb}^{1}, t^1_{\gga\gd}]_*=0\,.
\ee
Specific dependence on $\vec\y$ in $C$ is not going to be
important in what follows, so we just set $C=C(y, \vec\y)$ and,
correspondingly,
\be
t^1_{\al\gb}=-z_{\al}z_{\gb}\int_{0}^{1}\dr\tau\,\tau(1-\tau)\,C(-\tau
z; \vec\y)\,e^{i\tau zy}\,.
\ee
We proceed with the Taylor representation of field $C$,
\eqref{Cexp}
\be
C(y, \vec\y)\equiv e^{-iy \pp}\, C(y',
\vec\y)\Big|_{y'=0}\,,\qquad \pp_{\al}=-i\ff{\p}{\p y'^{\al}}\,.
\ee
Using \eqref{limst} it is easy to derive the following useful
product
\begin{align}
&z_{\al}z_{\gb}\,\int_{0}^{1}\dr\tau_1\,\tau_1(1-\tau_1)\,e^{i\tau_1z(y+\pp_1)}*
z_{\gga}z_{\gd}\,\int_{0}^{1}\dr\tau_2\,\tau_2(1-\tau_2)\,e^{i\tau_1z(y+\pp_2)}=\\
&z_{\al}z_{\gb}z_{\gga}z_{\gd}\int_{[0,1]^2}\dr\tau\dr\gs\,\gs(1-\gs)\tau^3(1-\tau)\,e^{i\tau
z(y+\gs \pp_1+(1-\gs)\pp_2)}\,.
\end{align}
Having it we obtain
\be\label{tt}
t^1_{\al\gb}*t^{1}_{\gga\gd}=z_{\al}z_{\gb}z_{\gga}z_{\gd}\int_{[0,1]^2}\dr\tau\dr\gs\,\gs(1-\gs)\tau^3(1-\tau)
\,e^{i\tau zy}\,C(-\tau\gs\, z, \vec\y)\star C(-\tau(1-\gs)\,z,
\vec\y)\,,
\ee
where $\star$ acts on $\vec\y$ via \eqref{starvec}. Since measure
in \eqref{tt} is invariant under $\gs\to 1-\gs$ one trivially
obtains \eqref{prove}. This completes the proof of \eqref{1}.


\begin{thebibliography}{20}

\bibitem{Bekaert:2022poo}
X.~Bekaert, N.~Boulanger, A.~Campoleoni, M.~Chiodaroli,
D.~Francia, M.~Grigoriev, E.~Sezgin and E.~Skvortsov, ``Snowmass
White Paper: Higher Spin Gravity and Higher Spin Symmetry,''
[arXiv:2205.01567].

\bibitem{Fronsdal:1978rb}
C.~Fronsdal, ``Massless Fields with Integer Spin,'' Phys. Rev. D
\textbf{18}, 3624 (1978)

\bibitem{Fang:1978wz}
J.~Fang and C.~Fronsdal, ``Massless Fields with Half Integral
Spin,'' Phys. Rev. D \textbf{18}, 3630 (1978)

\bibitem{Fronsdal:1978vb}
C.~Fronsdal, ``Singletons and Massless, Integral Spin Fields on de
Sitter Space (Elementary Particles in a Curved Space. 7.,'' Phys.
Rev. D \textbf{20}, 848-856 (1979)

\bibitem{Weinberg:1964ew}
S.~Weinberg, ``Photons and Gravitons in  $S$-Matrix Theory:
Derivation of Charge Conservation and Equality of Gravitational
and Inertial Mass,'' Phys. Rev. \textbf{135}, B1049-B1056 (1964)

\bibitem{Coleman:1967ad}
S.~R.~Coleman and J.~Mandula, ``All Possible Symmetries of the S
Matrix,'' Phys. Rev. \textbf{159}, 1251-1256 (1967)

\bibitem{Aragone:1979hx}
C.~Aragone and S.~Deser, ``Consistency Problems of Hypergravity,''
Phys. Lett. B \textbf{86}, 161-163 (1979)

\bibitem{Metsaev:1991mt}
R.~R.~Metsaev, ``Poincare invariant dynamics of massless higher
spins: Fourth order analysis on mass shell,'' Mod. Phys. Lett. A
\textbf{6}, 359-367 (1991)

\bibitem{Bekaert:2010hw}
X.~Bekaert, N.~Boulanger and P.~Sundell, ``How higher-spin gravity
surpasses the spin two barrier: no-go theorems versus yes-go
examples,'' Rev. Mod. Phys. \textbf{84}, 987-1009 (2012)


\bibitem{Roiban:2017iqg}
R.~Roiban and A.~A.~Tseytlin, ``On four-point interactions in
massless higher spin theory in flat space,'' JHEP \textbf{04}, 139
(2017) [arXiv:1701.05773].

\bibitem{Taronna:2017wbx}
M.~Taronna, ``On the Non-Local Obstruction to Interacting Higher
Spins in Flat Space,'' JHEP \textbf{05}, 026 (2017)
[arXiv:1701.05772].

\bibitem{Ponomarev:2022ryp}
D.~Ponomarev, ``Towards higher-spin holography in flat space,''
JHEP \textbf{01}, 084 (2023) [arXiv:2210.04035].

\bibitem{Ponomarev:2022qkx}
D.~Ponomarev, ``Chiral higher-spin holography in flat space: the
Flato-Fronsdal theorem and lower-point functions,'' JHEP
\textbf{01}, 048 (2023) [arXiv:2210.04036].

\bibitem{Fradkin:1986ka}
E.~S.~Fradkin and M.~A.~Vasiliev, ``Candidate to the Role of
Higher Spin Symmetry,'' Annals Phys. \textbf{177}, 63 (1987)


\bibitem{Fradkin:1987ks}
E.~S.~Fradkin and M.~A.~Vasiliev, ``On the Gravitational
Interaction of Massless Higher Spin Fields,'' Phys. Lett. B
\textbf{189}, 89-95 (1987)



\bibitem{Vasiliev:1989re}
M.~A.~Vasiliev, ``Higher Spin Algebras and Quantization on the
Sphere and Hyperboloid,'' Int. J. Mod. Phys. A \textbf{6},
1115-1135 (1991)

\bibitem{Vasiliev:1992av}
M.~A.~Vasiliev, ``More on equations of motion for interacting
massless fields of all spins in (3+1)-dimensions,'' Phys. Lett. B
\textbf{285}, 225-234 (1992)

\bibitem{Zinoviev:2001dt}
Y.~M.~Zinoviev, ``On massive high spin particles in AdS,''
[arXiv:hep-th/0108192].


\bibitem{Buchbinder:2001bs}
I.~L.~Buchbinder, A.~Pashnev and M.~Tsulaia, ``Lagrangian
formulation of the massless higher integer spin fields in the AdS
background,'' Phys. Lett. B \textbf{523}, 338-346 (2001)
[arXiv:hep-th/0109067].

\bibitem{Francia:2002pt}
D.~Francia and A.~Sagnotti, ``On the geometry of higher spin gauge
fields,'' Class. Quant. Grav. \textbf{20}, S473-S486 (2003)
[arXiv:hep-th/0212185].

\bibitem{Klebanov:2002ja}
I.~R.~Klebanov and A.~M.~Polyakov,
Phys. Lett. B \textbf{550}, 213-219 (2002) [arXiv:hep-th/0210114].

\bibitem{Leigh:2003gk}
R.~G.~Leigh and A.~C.~Petkou, ``Holography of the N=1 higher spin
theory on AdS(4),'' JHEP \textbf{06}, 011 (2003)
[arXiv:hep-th/0304217].

\bibitem{Sezgin:2003pt}
E.~Sezgin and P.~Sundell, ``Holography in 4D (super) higher spin
theories and a test via cubic scalar couplings,'' JHEP
\textbf{07}, 044 (2005) [arXiv:hep-th/0305040].

\bibitem{Vasiliev:2003ev}
M.~A.~Vasiliev, ``Nonlinear equations for symmetric massless
higher spin fields in (A)dS(d),'' Phys. Lett. B \textbf{567},
139-151 (2003) [arXiv:hep-th/0304049].

\bibitem{Metsaev:2005ar}
R.~R.~Metsaev, ``Cubic interaction vertices of massive and
massless higher spin fields,'' Nucl. Phys. B \textbf{759}, 147-201
(2006) [arXiv:hep-th/0512342].

\bibitem{Campoleoni:2010zq}
A.~Campoleoni, S.~Fredenhagen, S.~Pfenninger and S.~Theisen,
``Asymptotic symmetries of three-dimensional gravity coupled to
higher-spin fields,'' JHEP \textbf{11}, 007 (2010)
[arXiv:1008.4744].

\bibitem{Henneaux:2010xg}
M.~Henneaux and S.~J.~Rey, ``Nonlinear $W_{infinity}$ as
Asymptotic Symmetry of Three-Dimensional Higher Spin Anti-de
Sitter Gravity,'' JHEP \textbf{12}, 007 (2010) [arXiv:1008.4579].

\bibitem{Jevicki:2011ss}
A.~Jevicki, K.~Jin and Q.~Ye, ``Collective Dipole Model of AdS/CFT
and Higher Spin Gravity,'' J. Phys. A \textbf{44}, 465402 (2011)
[arXiv:1106.3983].

\bibitem{Iazeolla:2008ix}
C.~Iazeolla and P.~Sundell, ``A Fiber Approach to Harmonic
Analysis of Unfolded Higher-Spin Field Equations,'' JHEP
\textbf{10}, 022 (2008) doi:10.1088/1126-6708/2008/10/022
[arXiv:0806.1942].

\bibitem{Polyakov:2009pk}
D.~Polyakov, ``Interactions of Massless Higher Spin Fields From
String Theory,'' Phys. Rev. D \textbf{82}, 066005 (2010)
[arXiv:0910.5338].

\bibitem{Maldacena:2011jn}
J.~Maldacena and A.~Zhiboedov, ``Constraining Conformal Field
Theories with A Higher Spin Symmetry,'' J. Phys. A \textbf{46},
214011 (2013) [arXiv:1112.1016].

\bibitem{Maldacena:2012sf}
J.~Maldacena and A.~Zhiboedov, ``Constraining conformal field
theories with a slightly broken higher spin symmetry,'' Class.
Quant. Grav. \textbf{30}, 104003 (2013) [arXiv:1204.3882].


\bibitem{Vasiliev:2012vf}
M.~A.~Vasiliev, ``Holography, Unfolding and Higher-Spin Theory,''
J. Phys. A \textbf{46}, 214013 (2013) [arXiv:1203.5554].

\bibitem{Vasiliev:2015mka}
M.~A.~Vasiliev, ``Invariant Functionals in Higher-Spin Theory,''
Nucl. Phys. B \textbf{916}, 219-253 (2017) [arXiv:1504.07289].

\bibitem{Vasiliev:2018zer}
M.~A.~Vasiliev, ``From Coxeter Higher-Spin Theories to Strings and
Tensor Models,'' JHEP \textbf{08}, 051 (2018) [arXiv:1804.06520].


\bibitem{David:2020fea}
A.~David and Y.~Neiman, ``Bulk interactions and boundary dual of
higher-spin-charged particles,'' JHEP \textbf{03}, 264 (2021)
doi:10.1007/JHEP03(2021)264 [arXiv:2009.02893].




\bibitem{Giombi:2009wh}
S.~Giombi and X.~Yin, ``Higher Spin Gauge Theory and Holography:
The Three-Point Functions,'' JHEP \textbf{09}, 115 (2010)
arXiv:0912.3462].

\bibitem{Giombi:2013fka}
S.~Giombi and I.~R.~Klebanov, ``One Loop Tests of Higher Spin
AdS/CFT,'' JHEP \textbf{12}, 068 (2013) [arXiv:1308.2337].

\bibitem{Beccaria:2014xda}
M.~Beccaria and A.~A.~Tseytlin, ``Higher spins in AdS$_{5}$ at one
loop: vacuum energy, boundary conformal anomalies and AdS/CFT,''
JHEP \textbf{11}, 114 (2014) [arXiv:1410.3273].

\bibitem{Bae:2016rgm}
J.~B.~Bae, E.~Joung and S.~Lal, ``One-loop test of free SU(N )
adjoint model holography,'' JHEP \textbf{04}, 061 (2016)
[arXiv:1603.05387].

\bibitem{Sleight:2016dba}
C.~Sleight and M.~Taronna, ``Higher Spin Interactions from
Conformal Field Theory: The Complete Cubic Couplings,'' Phys. Rev.
Lett. \textbf{116}, no.18, 181602 (2016) [arXiv:1603.00022].

\bibitem{Sezgin:2017jgm}
E.~Sezgin, E.~D.~Skvortsov and Y.~Zhu, ``Chern-Simons Matter
Theories and Higher Spin Gravity,'' JHEP \textbf{07}, 133 (2017)
[arXiv:1705.03197].

\bibitem{Didenko:2017lsn}
V.~E.~Didenko and M.~A.~Vasiliev, ``Test of the local form of
higher-spin equations via AdS / CFT,'' Phys. Lett. B \textbf{775},
352-360 (2017) [arXiv:1705.03440].

\bibitem{Misuna:2017bjb}
N.~Misuna, ``On current contribution to Fronsdal equations,''
Phys. Lett. B \textbf{778}, 71-78 (2018) [arXiv:1706.04605].

\bibitem{Bekaert:2015tva}
X.~Bekaert, J.~Erdmenger, D.~Ponomarev and C.~Sleight, ``Quartic
AdS Interactions in Higher-Spin Gravity from Conformal Field
Theory,'' JHEP \textbf{11}, 149 (2015) [arXiv:1508.04292].

\bibitem{Sleight:2017pcz}
C.~Sleight and M.~Taronna, ``Higher-Spin Gauge Theories and Bulk
Locality,'' Phys. Rev. Lett. \textbf{121}, no.17, 171604 (2018)
[arXiv:1704.07859].

\bibitem{Boulanger:2015ova}
N.~Boulanger, P.~Kessel, E.~D.~Skvortsov and M.~Taronna, ``Higher
spin interactions in four-dimensions: Vasiliev versus Fronsdal,''
J. Phys. A \textbf{49}, no.9, 095402 (2016) [arXiv:1508.04139].

\bibitem{Vasiliev:2016xui}
M.~A.~Vasiliev, ``Current Interactions and Holography from the
0-Form Sector of Nonlinear Higher-Spin Equations,'' JHEP
\textbf{10}, 111 (2017) [arXiv:1605.02662].

\bibitem{Gelfond:2018vmi}
O.~A.~Gelfond and M.~A.~Vasiliev, ``Homotopy Operators and
Locality Theorems in Higher-Spin Equations,'' Phys. Lett. B
\textbf{786}, 180-188 (2018) [arXiv:1805.11941].

\bibitem{Didenko:2018fgx}
V.~E.~Didenko, O.~A.~Gelfond, A.~V.~Korybut and M.~A.~Vasiliev,
``Homotopy Properties and Lower-Order Vertices in Higher-Spin
Equations,'' J. Phys. A \textbf{51}, no.46, 465202 (2018)
[arXiv:1807.00001].

\bibitem{Didenko:2019xzz}
V.~E.~Didenko, O.~A.~Gelfond, A.~V.~Korybut and M.~A.~Vasiliev,
``Limiting Shifted Homotopy in Higher-Spin Theory and
Spin-Locality,'' JHEP \textbf{12}, 086 (2019) [arXiv:1909.04876].

\bibitem{Gelfond:2019tac}
O.~A.~Gelfond and M.~A.~Vasiliev, ``Spin-Locality of Higher-Spin
Theories and Star-Product Functional Classes,'' JHEP \textbf{03},
002 (2020) [arXiv:1910.00487].

\bibitem{Didenko:2020bxd}
V.~E.~Didenko, O.~A.~Gelfond, A.~V.~Korybut and M.~A.~Vasiliev,
``Spin-locality of $\eta^{2}$ and $ {\overline{\eta}}^2 $ quartic
higher-spin vertices,'' JHEP \textbf{12}, 184 (2020)
[arXiv:2009.02811].

\bibitem{Gelfond:2021two}
O.~A.~Gelfond and A.~V.~Korybut, ``Manifest form of the spin-local
higher-spin vertex $\varUpsilon ^{\eta \eta }_{\omega CCC}$,''
Eur. Phys. J. C \textbf{81}, no.7, 605 (2021) [arXiv:2101.01683].

\bibitem{Vasiliev:2022med}
M.~A.~Vasiliev, ``Projectively-compact spinor vertices and
space-time spin-locality in higher-spin theory,'' Phys. Lett. B
\textbf{834}, 137401 (2022) [arXiv:2208.02004].

\bibitem{Tarusov:2022qpo}
A.~A.~Tarusov, K.~A.~Ushakov and M.~A.~Vasiliev, ``Shifted
Homotopy Analysis of the Linearized Higher-Spin Equations in
Arbitrary Higher-Spin Background,'' [arXiv:2212.01908].

\bibitem{Didenko:2022eso}
V.~E.~Didenko and A.~V.~Korybut, ``On $z$-dominance, shift
symmetry and spin locality in higher-spin theory,''
[arXiv:2212.05006].

\bibitem{Sharapov:2022awp}
A.~Sharapov and E.~Skvortsov, ``Chiral higher spin gravity in
(A)dS4 and secrets of Chern\textendash{}Simons matter theories,''
Nucl. Phys. B \textbf{985}, 115982 (2022) [arXiv:2205.15293].

\bibitem{Sharapov:2022nps}
A.~Sharapov, E.~Skvortsov, A.~Sukhanov and R.~Van Dongen, ``More
on Chiral Higher Spin Gravity and Convex Geometry,''
[arXiv:2209.15441].

\bibitem{Didenko:2022qga}
V.~E.~Didenko, ``On holomorphic sector of higher-spin theory,''
JHEP \textbf{10}, 191 (2022) [arXiv:2209.01966].

\bibitem{Korybut:2022kdx}
A.~V.~Korybut, A.~A.~Sevostyanova, M.~A.~Vasiliev and
V.~A.~Vereitin, ``Disentanglement of Topological and Dynamical
Fields in 3d Higher-Spin Theory within Shifted Homotopy
Approach,'' [arXiv:2211.15778].

\bibitem{Eastwood:2002su}
M.~G.~Eastwood, ``Higher symmetries of the Laplacian,'' Annals
Math. \textbf{161}, 1645-1665 (2005) [arXiv:hep-th/0206233
[hep-th]].

\bibitem{Vasiliev:1988sa}
M.~A.~Vasiliev, ``Consistent Equations for Interacting Massless
Fields of All Spins in the First Order in Curvatures,'' Annals
Phys. \textbf{190} (1989), 59-106

\bibitem{Misuna:2019ijn}
N.~Misuna, ``On unfolded off-shell formulation for higher-spin
theory,'' Phys. Lett. B \textbf{798}, 134956 (2019)
[arXiv:1905.06925].

\bibitem{Misuna:2022cma}
N.~G.~Misuna, ``Unfolded Dynamics Approach and Quantum Field
Theory,'' [arXiv:2208.04306].

\bibitem{Vasiliev:2004cm}
M.~A.~Vasiliev, ``Higher spin superalgebras in any dimension and
their representations,'' JHEP \textbf{12}, 046 (2004)
[arXiv:hep-th/0404124].

\bibitem{Korybut_in_prep}
A.~V.~Korybut, ``On consistency of the interacting
(anti)holomorphic higher-spin sector,'' to be published


\bibitem{Sagnotti:2004pod}
A.~Sagnotti, E.~Sezgin and P.~Sundell, ``On higher spins with a
strong Sp(2,R) condition,'' [arXiv:hep-th/0501156].


\bibitem{Bekaert:2004qos}
X.~Bekaert, S.~Cnockaert, C.~Iazeolla and M.~A.~Vasiliev,
``Nonlinear higher spin theories in various dimensions,''
[arXiv:hep-th/0503128].

\bibitem{Joung:2014qya}
E.~Joung and K.~Mkrtchyan, ``Notes on higher-spin algebras:
minimal representations and structure constants,'' JHEP
\textbf{05}, 103 (2014) [arXiv:1401.7977].

\bibitem{Alkalaev:2019xuv}
K.~Alkalaev and X.~Bekaert, ``Towards higher-spin AdS$_2$/CFT$_1$
holography,'' JHEP \textbf{04}, 206 (2020)
doi:10.1007/JHEP04(2020)206 [arXiv:1911.13212].

\bibitem{Prokushkin:1998bq}
S.~F.~Prokushkin and M.~A.~Vasiliev, ``Higher spin gauge
interactions for massive matter fields in 3-D AdS space-time,''
Nucl. Phys. B \textbf{545}, 385 (1999) [arXiv:hep-th/9806236].

\bibitem{DeFilippi:2019jqq}
D.~De Filippi, C.~Iazeolla and P.~Sundell, ``Fronsdal fields from
gauge functions in Vasiliev\textquoteright{}s higher spin
gravity,'' JHEP \textbf{10}, 215 (2019) [arXiv:1905.06325].

\bibitem{Iazeolla:2020jee}
C.~Iazeolla, ``On boundary conditions and spacetime/fibre duality
in Vasiliev's higher-spin gravity,'' PoS \textbf{CORFU2019}, 181
(2020) [arXiv:2004.14903].

\bibitem{DeFilippi:2021xon}
D.~De Filippi, C.~Iazeolla and P.~Sundell, ``Metaplectic
representation and ordering (in)dependence in
Vasiliev\textquoteright{}s higher spin gravity,'' JHEP
\textbf{07}, 003 (2022) [arXiv:2111.09288].


\end{thebibliography}
\end{document}